\documentclass[twocolumn, 10pt]{IEEEtran}

\usepackage{enumitem}
\usepackage{hyperref}
\usepackage{breakurl}

\usepackage{hyperref} 
\usepackage{graphicx}
\usepackage{amssymb}
\usepackage{amsmath}
\usepackage{mathtools}
\usepackage{cite}
\usepackage{stfloats}
\usepackage{psfrag}
\usepackage[mathscr]{euscript}
\usepackage{acronym}  
\usepackage{booktabs} 
\usepackage[table]{xcolor}
\usepackage{graphicx}

\usepackage{caption}
\usepackage{subfig}

\DeclareMathAlphabet{\mathrmbf}{\encodingdefault}{\rmdefault}{bx}{n}

\newcommand{\mcal}{\mathcal}
\newcommand{\mbb}{\mathbb}
\newcommand{\mbf}{\mathbf}

\newcommand{\msf}{\mathsf}

\newtheorem{theorem}{Theorem}
\newtheorem{lemma}{Lemma}
\newtheorem{definition}{Definition}

\def\UE{\mathrm{UE}}

\def\Hf{\mathsf{H}}
\def\Ef{\mathsf{E}_{\text{S}}}

\def\ASR{\mathcal{R}}
\def\ID{\msf{ID}}
\def\AID{\msf{AID}}
\def\GID{\msf{GID}}
\def\UID{\msf{UID}}
\def\TID{\msf{TID}}

\usepackage{array}
\newcolumntype{P}[1]{>{\vspace*{0.07cm}\centering\arraybackslash}p{#1}}
\newcolumntype{L}[1]{>{\vspace*{0.07cm}\raggedright\arraybackslash}m{#1}}
\newcolumntype{M}[1]{>{\vspace*{0.07cm}\centering\arraybackslash}m{#1}}

\usepackage{pbox}

\begin{document}
%

\title{{GRAAD: Group Anonymous and Accountable D2D Communication in Mobile Networks}}
%
%
%
%
%

\author{	
	\vspace{0.2cm}
	Ruei-Hau~Hsu, \textit{Member, IEEE}, 
        Jemin~Lee, \textit{Member, IEEE}, 
        Tony~Q.~S.~Quek, \textit{Senior Member, IEEE}, and
        Jyh-Cheng~Chen, \textit{Fellow, IEEE}
\IEEEcompsocitemizethanks{\IEEEcompsocthanksitem R.-H. Hsu is with iTrust, Centre for Research in Cyber Security, Singapore University of Technology and Design, Singapore, 487372
(E-mail: richard\_hsu@sutd.edu.sg). 
J.~Lee is with Department of Information and Communication Engineering, Daegu Gyeongbuk Institute of Science and Technology, Korea, 43016 
(Email: jmnlee@dgist.ac.kr). 
T.~Q.~S. Quek is with Information Systems Technology and Design Pillar, Singapore University of Technology and Design, Singapore, 487372
(Email: tonyquek@sutd.edu.sg).
J.-C. Chen is with Department of Computer Science, National Chiao Tung University, Taiwan, 300.
(Email: jcchen@ieee.org)
The contact author is J. Lee.
}
}

\IEEEcompsoctitleabstractindextext{%
\begin{abstract}
Device-to-Device~(D2D) communication is mainly launched by the transmission requirements between devices 
for specific applications such as Proximity Services in Long-Term Evolution Advanced~(LTE-A) networks, and each application will form a group of registered devices for the network-covered and network-absent D2D communications.
%
During the applications of D2D communication, each device needs to identify the other devices of the same group in proximity by their group identity. This leads to the exposure of group information, by which the usage of applications can be analyzed by eavesdroppers.
%
Hence, this work introduces network-covered and network-absent authenticated key exchange protocols for D2D communications
to guarantee accountable group anonymity, end-to-end security to network operators, as well as traceability and revocability for {accounting} and management requirements. 
We formally prove the security of those protocols, and also develop an analytic model to evaluate the quality of authentication protocols by authentication success rate in D2D communications.
Besides, we implement the proposed protocols on android mobile devices to evaluate the computation costs of the protocols. {We also evaluate the authentication success rate by the proposed analytic model and prove the correctness of the analytic model via simulation.}  
Those evaluations show that the proposed protocols are feasible to the performance requirements of D2D communications.


\end{abstract}

}
\maketitle

\IEEEdisplaynotcompsoctitleabstractindextext

\IEEEpeerreviewmaketitle

\vspace{-1mm}
\section{Introduction}
\IEEEPARstart{D}{ue} to the dramatic growth of the number of mobile devices, providing mobile communication services with higher throughput, lower traffic overhead, and lower energy consumption are challenges. Although LTE-A physical-layer provides even higher communication capability~\cite{3gpp_PHY_des}, the resource allocation in the evolved universal terrestrial radio access network~(E-UTRAN) to high density mobile devices remains dilemma when the resource is limited. The 3rd generation partnership project (3GPP) {proposes} D2D communication service in LTE-A, called Proximity Service (ProSe)~\cite{3gpp_ProSe_FeaStudy,3gpp_ProSe_Study_Arch} {with} three main purposes as follows: 1) the mobile network operator can offload traffic of E-UTRAN and Evolved Packet System~(EPS)~\cite{3gpp_Sys-Arch}, which is the core network~(CN) of the LTE-A system; 2) D2D communication may support social network service, information sharing, advertising, gaming, and conferencing services; and 3) the high availability of D2D communication can be used to support public safety services. {Besides, security is essential to support the correctness of the functions and the availability for D2D communications.} 

{In ProSe, D2D communication can be classified as the {\it network-covered} and {\it network-absent} 
according to whether its control components are connected to CN (covered) or not (absent).} 
The authenticated key exchange~(AKE) in ProSe have to consider the connectivity between user equipments~(UEs) and CN and should provide security protection from various kinds of attacks. Certain security threats have been discussed in~\cite{D2D_LTE_AYRA}, i.e., eavesdropping between UEs, impersonation attack on UE or evolved NodeB~(eNB), and active attack {by injecting messages into traffic data or control data.} 

{AKE guarantees} the identification {by mutual authentication} and confidentiality of communication {by key exchange} in computer networks~\cite{ToN_Sec_BGHJKMY95,ToN_Sec_LLY06,ToN_Sec_FHH10}. Additionally, the anonymous protection {to {\it user identity}} is critical due to the broadcast nature of wireless communications. This security requirement has been carefully deliberated in~\cite{Wireless_Auth_ZM04,Wireless_Auth_JLSS06,Wireless_Auth_TO08,Wireless_Auth_YHWD10,Wireless_Auth_HBCCY11,Wireless_Auth_HCCB12,Wireless_Auth_RH13,Wireless_Auth_LLLLZS14,Wireless_Auth_GH15,Wireless_Auth_HCG15}. In mobile networks, an UE should complete authentication for identity identification in advance of requesting for services when roaming to a foreign network~(FN). The user anonymous authentication prevents eavesdroppers or/and FN from disclosing the real identities of UEs in every authentication session, whereby the locations of UEs~(i.e., footprints) may be tracked. 

{Anonymity} can be divided as two levels, partial user anonymity and full user anonymity. Partial user anonymous authentication conceals identities from eavesdroppers, excluding FNs~\cite{Wireless_Auth_ZM04,Wireless_Auth_JLSS06,Wireless_Auth_TO08} and full user anonymous authentication additionally considers FNs as eavesdroppers ~\cite{Wireless_Auth_YHWD10,Wireless_Auth_HBCCY11,Wireless_Auth_HCCB12,Wireless_Auth_RH13,Wireless_Auth_LLLLZS14,Wireless_Auth_HCG15}. In case of full user anonymity, traceability and revocability are essential to support the permitted network operators to trace and revoke user identities for management purposes. Certain traceability and revocability techniques~\cite{Wireless_Auth_YHWD10,Wireless_Auth_HBCCY11,Wireless_Auth_HCG15} have been introduced to cancel the anonymity protection in secure wireless communications.

The aforementioned studies provide elegant solutions to support anonymous and secure wireless communication between users and networks. For D2D communications, two secure D2D communication systems~\cite{TVT_D2D_ZCHQ16,TIFS_D2D_ZWYL16} are proposed to support data sharing with distinct application scenarios. One~\cite{TVT_D2D_ZCHQ16} supports pseudonymity protection, where each real identity is replaced with a corresponding pseudo identity so that the sessions from the same device are traceable. The other~\cite{TIFS_D2D_ZWYL16} offers partial user anonymity, where system is able to trace the footprints of users. 

{Nevertheless}, the new anonymity issue for the group information arises when direct communications between devices are launched for specific applications. In ProSe-enabled devices, the group is formed by devices using the same application. 
During the establishment of a D2D communication including device discovery procedure, the device, which initiates the D2D communication, needs to announce messages with an application identity so that it can be discovered by other devices in the same application group. 
Attackers may collect and analyze the application usage information and launch distributed denial-of-service (DDoS) attacks to specific groups or observe the behaviors of users in proximity for malicious purposes.
Hence, {\it group anonymous protection with traceability} should be considered in D2D communication to protect the application information included in the announcing message need to be protected while establishing D2D communications.

End-to-end security is another required security property as devices exchange messages via D2D communications. This security property prevents system operators, who help to establish D2D communications, from eavesdropping exchanging messages between devices.


{\bf Security Difference.} Compared to wireless communication, D2D communications additionally consider entity authentication without involving security infrastructure~(i.e., authentication server), privacy protection against network infrastructure, group anonymity preventing from exposing group or application related identity, and end-to-end security among devices.


\subsection{Contributions}
This work presents two group anonymous {authenticated} key exchange protocols for network-covered and network-absent D2D communications {in mobile networks} to support {identity and} group (application) anonymity, {accountability~(i.e., }traceability {and} revocability), and end-to-end security {against insider attacker}~(between devices).
Specifically, we first propose the group-anonymous D2D communication with CN-assistance (CN-GD2C) protocol adopting {identity-based encryption~(IBE) against chosen ciphertext attacks~(IND-CCA)}, Diffie-Hellman key exchange, symmetry-based encryption, and hash functions. We also propose the group-anonymous {authenticated} key exchange for network-absent D2D communication (NA-GD2C) protocol by utilizing { the new proposed identity-based $k$-anonymity secret handshake scheme} with {the \textit{encryptions and proof technique} by combining public-key encryptions~(key-private encryption and Linear encryption) and zero-knowledge proof in the design}.
{We then formally prove the security of these two protocols and develop an analytic model using queueing theory to evaluate authentication success rates of the proposed protocols and demonstrate the scalability and efficiency of the proposed protocols. 
We also implement the proposed protocols to estimate the computation costs on mobile devices, 
and obtain the authentication success rates by both simulation and analytic model.} 

{The remainder of this paper is organized as follows. In Section~\ref{sec:system_security_model}, we introduce the system and security models of D2D communication in mobile networks. In Section~\ref{sec:GD2C}, we propose group anonymous D2D communication protocols of network-covered and network-absent cases. The security analysis and performance evaluation on the proposed protocols are presented in Section~\ref{sec:security_analysis} and Section~\ref{sec:performance_evaluation}, respectively. Finally, we conclude this work in Section~\ref{sec:conclusions}.}

\section{System and Security Models}\label{sec:system_security_model}
This section introduces the system model including the functions of the system components and the operating procedures for secure D2D communications in ProSe~\cite{3gpp_ProSe_Stage2,3gpp_ProSe_Security}. 
We then introduce the behaviors of the attackers, and propose the security {model and its definitions} of the proposed system. 
\subsection{{System Model and Security Requirements}}\label{subsec:sys_sec_model}
\subsubsection{{The System Model}}\label{subsubsec:sys_model}
In ProSe, there are two kinds of UEs, announcing UE~(A-UE), who requests for establishing D2D communication, and monitoring UE~(M-UE), who monitors the requests of D2D communications in proximity. In network-covered D2D communications, each UE shares long-term secret with authentication center/home subscriber server~(AuC/HSS), which is response for the user subscription management, user authentication and session key management. 
When UEs establish D2D communication for ProSe, they can be either inside or outside of the coverage of CN. Each UE can access E-UTRAN via eNB in its coverage. 
{Before establishing D2D communication, each A-UE and M-UE need to register to ProSe function to obtain the required parameters for D2D communication configuration. 

In network-covered D2D communications, both A-UE and M-UE attach to CN for device discovery procedure with authentication and authorization. During a device discovery, the ProSe function will send authentication requests to AuC/HSS to authenticate participant devices. Once authentication requests are received, HSS/AuC will produce the corresponding authentication token for ProSe function to authenticate the UEs. By the aforesaid parameters from ProSe function, the A-UE can announce device discovery messages and be discovered by the M-UEs of the same application groups.}

In network-absent D2D communications, HSS/AuC assigns each UE an identity with the corresponding private key of an IBE system for secure D2D communication. The identity is valid for a pre-defined duration and can be revoked as required. In both communication modes, the application identity of each UE for services are maintained by ProSe function.
\subsubsection{The Security Requirements} 
{According to the system model, we analyze and propose the following security requirements that are urgently required in D2D communications.}
\begin{itemize}[leftmargin=*]
  \item{\it {Authenticated key exchanged with end-to-end security:}}
  This is to guarantee the authentication of intended participants and confidentiality of the transmission between two UEs in ProSe. {Typically, two parties achieve authenticated key exchange~(AKE) with end-to-end secure communication based on a long-term secret only known by them. However, UEs only share long-term secret key with the HSS/AuC located in CN. Hence, AKE between two UEs needs the participation of the HSS/AuC and this leads the exchanged session key between two UEs will be known by the HSS/AuC. Namely, the communication between two UEs will be exposed to the CN. In the sense of D2D communications, communication confidentiality between two UEs should be guaranteed.} {Hence, ProSe needs to achieve {\it authenticated key exchange with end-to-end security} between two UEs.}
  \item{\it {Identity and} group anonymity:}
  {Identity anonymity is to guarantee that the identities of the participants in each AKE session are protected and cannot be linked between sessions to prevent outsider attackers from tracing the footprints of the participants. Besides that, group anonymity is also essential in ProSe since the usage of applications of UEs in ProSe is considered as sensitive information, which may be analyzed and utilized for disturbing services.} Only two UEs {in the same application group} can successfully authenticate each other and exchange a session key. If two UEs belong to different groups, they are unable to learn identity and group information of each other in authentication~\cite{SH_BDSS03}. {Obviously, AKE with group anonymity also guarantees identity anonymity.}
  \item{\it Traceability and revocability:}
  Traceability in group anonymous authentication 
  guarantees the identity and group information of participants in every successful authentication session can be disclosed when {the identities of UEs in ProSe are required for management or accounting purposes by CN.} Only authorized {entity}, e.g., HSS/AuC or ProSe, can disclose the group and identity information. Revocability ensures the identity of every UE can be revoked to terminate D2D services.
\end{itemize}

\begin{figure}[!t]
\begin{center}
\includegraphics[width=1\columnwidth]
{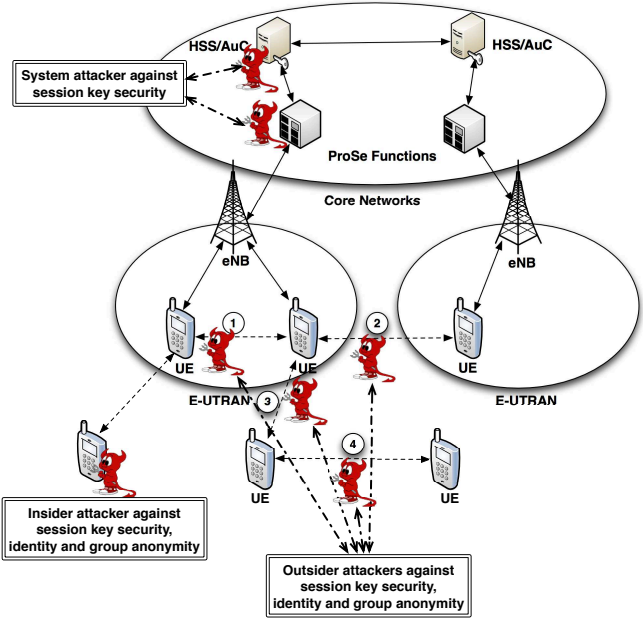} 
\caption{Attacker Models of ProSe between UEs 
}
\label{fig:prose_attacker}
\end{center}
\end{figure}

\subsection{{Security Models and Definitions}}\label{subsec:sec_model_def}
{Figure~\ref{fig:prose_attacker} shows the attacker model of ProSe. An outsider attacker may eavesdrop the communications including the exchanged messages, the identity information, or the group information, between two UEs. It may break the confidentiality of communications, trace the footprints of UEs, or probe the using applications according to the group information. The outsider attacker may also impersonate as a legal user to pass the authentication and exchange a common session key with any legal user. Furthermore, a legal user can be an attacker to achieve mutual authentication and exchange a common session key with any user belonging to different group. Additionally, the ProSe function or HSS/AuC can be system attackers, who eavesdrop the exchanged messages between devices in D2D communications. This kind of attacker is commonly not considered by the security solutions in mobile networks. However, it makes sense to consider secure communication against the system attackers as the messages are merely exchanged between devices in D2D communications. Before defining the attackers, we define an authenticated key exchange protocol and the capabilities of attackers in the protocol as follows.}

{The proposed protocol is $\Pi$, and $\Pi_{U,V}^{t_1}$ and $\Pi_{V,U}^{t_2}$ are regarded as two instances to model two users $U$ and $V$ being the partners of each other in the communication session $t_1$ and $t_2$ of $\Pi$. We say that a matching conversation involving $\Pi_{U,V}^{t_1}$ and $\Pi_{V,U}^{t,2}$ if and only if $t_1=t_2$ and $U$ and $V$ are partners. The capability of an attacker can be captured by the following oracles.
\begin{itemize}[leftmargin=*]
\item $\textbf{Execute}(\Pi_{U,V}^{t_1},\Pi_{V,U}^{t_2}):$ This oracle models a passive attacker, who can intercept all communications between $\Pi_{U,V}^{t_1}$ and $\Pi_{V,U}^{t_2}$.
\item $\textbf{Send}(\Pi_{U,V}^{t_1},m):$ This oracle models an active attacker, who sends a message $m$ to $\Pi_{U,V}^{t_1}$.
\item $\textbf{Reveal}(\Pi_{U,V}^{t_1}):$ This oracle models the exposure of the accepted session key of $U$ shared with its partner $V$ in the session $t_1$.
\item $\textbf{Corrupt}(\Pi_{U,V}^{t_1}):$ This oracle models the exposure of the long-term secret key of $U$ during the session $t_1$ with its partner $V$.
\item $\textbf{Test}(\Pi_{U,V}^{t_1}):$ When an attacker queries this oracle, it will return a real session key, accepted by $U$ with its partner $V$ in the session $t_1$, or a random string according to a random bit if the negotiation of the session key is complete. The query of this oracle is failed, if the session key is not negotiated.
\item $\textbf{TestID}(\Pi_{U,V}^{t_1}):$ When an attacker queries this oracle, it will return the real identity of $U$ or a random string according to a random bit when $U$ and $V$ are accepted each other with a negotiated session key. The query of this oracle is failed, if the AKE is not fulfilled between $U$ and $V$.
\item $\textbf{TestGroup}(\Pi_{U,V}^{t_1}):$ When an attacker queries this oracle, it will return the group information of $U$ or a random string according to a random bit when $U$ and $V$ are accepted each other with a negotiated session key. The query of this oracle is failed, if the AKE is not fulfilled between $U$ and $V$.
\end{itemize}}

{We then define the security of AKE in ProSe according to the discussed security requirements as follows.
\begin{definition}[Mutul Authentication]\label{def:mutual_auth}
There are a simulator $\mcal{S}$, who simulates $\Pi_{U,V}^{t_1}$ or $\Pi_{V,U}^{t_2}$ by $\Pi$, and a probabilistic polynomial-time~(PPT) attacker $\mcal{A}$ of $\Pi$, who can query \textbf{Execute} and \textbf{Send} in polynomial time. After oracle queries, $\mcal{A}$ sends a message to be accepted by $\Pi_{U,V}^{t_1}$ or $\Pi_{V,U}^{t_2}$ with the advantage as follows:
\begin{equation}
\textbf{Adv}_{\mcal{A}}^{\textrm{MA}} = \Pr[\mcal{A}\textrm{ accepted by }\Pi_{U,V}^{t_1}\textrm{ or }\Pi_{V,U}^{t,2}]
\end{equation}
\end{definition}
The mutual authentication security of $\Pi$ is guaranteed for $U$ and $V$ if 1) $\Pi_{U,V}^{t_1}$ and $\Pi_{V,U}^{t_2}$ has a matching conversation with and are accepted by each other and 2) $\textbf{Adv}_{\mcal{A}}^{\textrm{MA}}$ is negligible.
\begin{definition}[Key Exchange against System Operator]\label{def:session_key_sec}
$\mcal{S}$ simulates $\Pi_{U,V}^{t_1}$ and $\Pi_{U,V}^{t_2}$ by $\Pi$ to interact with $\mcal{A}$, who is either an outsider attacker or a system attacker~(i.e., HSS/AuC or ProSe function), and can query \textbf{Execute} and \textbf{Send} in polynomial time. After oracle queries, $\Pi_{U,V}^{t_1}$ and $\Pi_{V,U}^{t_2}$ are accepted by each other with an exchanged session key $K_{U,V}$, $\mcal{A}$ queries $\textbf{Test}$ to obtain $K_{U,V}$ or a random string from $\Pi_{U,V}^{t_1}$ or $\Pi_{V,U}^{t_2}$ according to a random bit $b\in\{0,1\}$. Then, $\mcal{A}$ outputs a guess $b'\in\{0,1\}$ with the following advantage.
\begin{equation}
\textbf{Adv}_{\mcal{A}}^{\textrm{SK}}=\left\{\Pr[b=b'] - 1/2\right\}
\end{equation}
If $\textbf{Adv}_{\mcal{A}}^{\textrm{SK}}$ is negligible, we say $\Pi$ achieves session key security.
\end{definition}
\begin{definition}[Identity Anonymity]\label{def:id_anon}
$\mcal{S}$ simulates $\Pi_{U,V}^{t_1}$ and $\Pi_{U,V}^{t_2}$ by $\Pi$ interacting with $\mcal{A}$ and $\mcal{A}$ can query \textbf{Execute} and \textbf{Send} in polynomial time. After oracle queries, $\Pi_{U,V}^{t_1}$ and $\Pi_{V,U}^{t_2}$ are accepted by each other, $\mcal{A}$ queries $\textbf{TestID}$ to obtain $\ID_{U}$ or $\ID_{V}$, or a random string from $\Pi_{U,V}^{t_1}$ or $\Pi_{V,U}^{t_2}$ according to a random bit $b\in\{0,1\}$. Then, $\mcal{A}$ outputs a guess $b'\in\{0,1\}$ with the following advantage.
\begin{equation}
\textbf{Adv}_{\mcal{A}}^{\textrm{anon\_id}}=\left\{\Pr[b=b'] - 1/2\right\}
\end{equation}
If $\textbf{Adv}_{\mcal{A}}^{\textrm{anon\_id}}$ is negligible, we say $\Pi$ achieves identity anonymity.
\end{definition}\label{def:group_anon}
\begin{definition}[Group Anonymity]\label{group_anon}
The security of group anonymity is similar to that of identity anonymity. Instead, $\mcal{A}$ queries $\textrm{TestGroup}$ to obtain the group information or a random string from $\Pi_{U,V}^{t_1}$ or $\Pi_{V,U}^{t,2}$ according to $b\in\{0,1\}$. The advantage of guessing $b$ from the output $b'\in\{0,1\}$ of $\mcal{A}$ is as follows.
\begin{equation}
\textbf{Adv}_{\mcal{A}}^{\textrm{anon\_group}}=\left\{\Pr[b=b'] - 1/2\right\}
\end{equation}
If $\textbf{Adv}_{\mcal{A}}^{\textrm{anon\_group}}$ is negligible, we say $\Pi$ achieves group anonymity.
\end{definition}
}

%

\section{Proposed Group Anonymous D2D Communications}\label{sec:GD2C}

\begin{table}[t!]
\caption{Notations}
\label{table:notations}
\begin{center}
{\footnotesize 
\begin{tabular}{M{2.2cm}|L{5.9cm}} \hline\hline
Notation & Description\\\hline\hline
\vspace{-.5mm}$\textrm{UE}_i$ & \vspace{-.5mm}user equipments $i$\\
\vspace{-.5mm}$\textrm{eNB}$ & \vspace{-.5mm} evolved node B\\
\vspace{-.5mm}HSS/AuC & \vspace{-.5mm}home subscriber server/authentication center \\
\vspace{-.5mm}$\ID_i$ & \vspace{-.5mm}the real identities of $\textrm{UE}_i$\\
\vspace{-.5mm}$\UID_i$ & \vspace{-.5mm}the identity in IBE system\\
\vspace{-.5mm}$\AID_i$ & \vspace{-.5mm}the application identity of $\textrm{UE}_i$\\
\vspace{-.5mm}$\GID_i$ & \vspace{-.5mm}the group identity of $\textrm{UE}_i$\\
\vspace{-.5mm}$K_i$ & \vspace{-.5mm}the shared secret key between $\textrm{UE}_i$ and HSS/AuC \\
\vspace{-.5mm}$K_{\text{p}}$ & \vspace{-.5mm}the secret key only known by ProSe Function \\
\vspace{-.5mm}$AK_i$ & \vspace{-.5mm}the authorization key of ProSe function for $\textrm{UE}_i$\\
\vspace{-.5mm}$\msf{param}$  & \vspace{-.5mm}the public parameters of an ID-based encryption\\
 \vspace{-.5mm}        &\vspace{-.5mm} system \\
\vspace{-.5mm}$d_{\msf{ID}}$ &\vspace{-.5mm} the IBE private key corresponding to $\msf{ID}$\\
\vspace{-.5mm}$\mathsf{E}_{\textrm{IC}}(pk,ID,x)$\vspace{-.5mm} & IND-CCA secure ID-based encryption with the\\ 
\vspace{-.5mm}&\vspace{-.5mm} system public key $pk$, the identity $ID$, and the input $x$ \\
 \vspace{-.5mm}$\mathsf{E}_{\textrm{S}}(K,x_1,x2,\cdots)$& \vspace{-.5mm}symmetric Encryption with inputs $x_1,x_2,$ etc.\\
 \vspace{-.5mm}$\mcal{X}=g^{x},\mcal{Y}=g^{y}$  & \vspace{-.5mm}a Diffie-Hellman tuple~\cite{DDH_Boneh}, where $(x,y)\in{}\mbb{Z}_{p}^{2}$\\
 \vspace{-.5mm}&\vspace{-.5mm} and $g\in\mbb{G}$.\\
 \vspace{-.5mm}$\msf{H}(.)$, $f_{0}(.)$, $\msf{H}_1(.)$,$\msf{H}_2(.)$  & \vspace{-.5mm}secure one-way hash functions, where  $\Hf:\{0,1\}^{*}\rightarrow{}\{0,1\}^{l}$, $f_{0}:\mbb{Z}_p^{3}\rightarrow{}\{0,1\}^{l}$,\\
 \vspace{-.5mm}&\vspace{-.5mm}$\msf{H}_1:\{0,1\}^{*}\rightarrow{}\mbb{G}^{*}$, $\msf{H}_2:\mbb{G}_T\rightarrow{}\{0,1\}^{n}$ \\
 \hline\hline 
\end{tabular}}
\end{center}
\end{table}

In this section, we propose two group anonymous schemes including the main building blocks for network-covered and network-absent cases.
\vspace*{-3mm}
{\subsection{Preliminary}
This section introduces the preliminaries of the proposed protocols.
\subsubsection{Bilinear Groups}
We first define the used bilinear map operation~\cite{GS_BBS04,BP_BLS01,Pairing_MNT01,Pairing_RS}. 
The bilinear maps is defined as $e:\mathbb{G}_1\times \mathbb{G}_2 \rightarrow \mathbb{G}_T$, 
where all group $\mbb{G}_1$, $\mbb{G}_2$, and $\mbb{G}_T$ are multiplicative and of prime order $p$. 
When $g_1$ is a generator of $\mbb{G}_1$ and $g_2$ is a generator of $\mbb{G}_2$, 
 there exists a computable isomorphism $\psi$ from $\mbb{G}_2$ to $\mbb{G}_1$ such as $\psi(g_1)=g_1$. 
The map $e$ has the following properties: 1) bilinearity: for all $u\in \mbb{G}_1$, $b\in \mbb{G}_2$, and $(a,b)\in\mbb{Z}_p^2$, $e(u^a,v^b)=e(u,v)^{ab}$; and 2) non-degeneracy: $e(g_1,g_2)\neq 1$.
\subsubsection{Identity-based Encryption}\label{subsubsec:IBE}
The concept of identity-based encryption~(IBE) is to eliminate the management costs of user certificates, i.e., verifying its correctness and its revocation. A usable IBE is first proposed by Boneh and Franklin~\cite{IBE_BF01}~(BF-IBE) with IND-CCA in random oracle model. It consists of four algorithms as follows.
\begin{itemize}[leftmargin=*]
\item{\bf Setup:} This algorithm is given a security parameter $\kappa$ to generate a prime $p$ and two bilinear groups $\mbb{G}$ and $\mbb{G}_T$ such that a bilinear map $e: \mbb{G}\times\mbb{G}\rightarrow{}G_T$ holds. It then chooses a random generator $g\in{}\mbb{G}$, sets $G_{pub}=g^{s}$ for a randomly selected $s\in\mbb{Z}_{p}$, and chooses two cryptographic hash functions, $\msf{H}_{1}:\{0,1\}^{*}\rightarrow{}\mbb{G}^{*}$ and $\msf{H}_2:\mbb{G}_T\rightarrow\{0,1\}^{n}$ for some $n$. The message space is $\mcal{M}=\{0,1\}^{n}$ and the ciphertext space is $\mcal{C}=\mbb{G}^{*}\times\{0,1\}^{n}$. The system parameters are $\msf{param}=\{p,\mbb{G},\mbb{G}_T,e,n,g,G_{pub},\msf{H}_1,\msf{H}_2\}$ and the master key of the system is $\msf{msk}=s$.
\item{\bf Extract:} This algorithm is given an identity $\msf{ID}\in\{0,1\}^{*}$ and computes the corresponding private key $d_{\msf{ID}}=Q_{\msf{ID}}^{s}$, where $Q_{\msf{ID}}=\msf{H}_1(\ID)\in\mbb{G}^{*}$.
\item{\bf Encrypt:} Given a message $M\in\mcal{M}$ and the identity $\msf{ID}$ as the public key, this algorithm encrypts the message as $C=\{g^{r},M\oplus{}\msf{H}_2(g_{\msf{ID}}^{r})\}=\{U,V\}$, where $g_{\msf{ID}}=e(Q_{\msf{ID}},G_{pub})\in\mbb{G}^{*}_{T}$ and $Q_{\msf{ID}}=\msf{H}_1(\msf{ID})\in\mbb{G}^{*}$.
\item{\bf Decrypt:} Given a ciphertext $C$ encrypted by $\msf{ID}$ and the private key $d_{\msf{ID}}$, the algorithm decrypts the message by $M=V\oplus{}\msf{H}_2(e(d_{\msf{ID}},U))=M$.
\end{itemize}
}

\noindent{{\bf Security of IBE.} The notion of ciphertext indistinguishability for the security of public-key encryption has been introduced to make an attacker obtain no information of the plaintext from a given ciphertext~\cite{EncSec_GM84}.
A stronger security notion of IND-CCA is proposed to satisfy the security requirement of secure communication~\cite{EncSec_CHK04}, where the attacker can decrypt any chosen ciphertexts other than the target ciphertext.}

{An IBE is IND-CCA secure if the advantage of any probabilistic polynomial-time~(PPT) adversary, $\mathcal{A}$, in the following game is negligible:
 1) $\mcal{A}$ issues $m$ queries for the private keys $d_{\msf{ID}_{i_1}},...,d_{\msf{ID}_{i_m}}$ of $\msf{ID}_{i_1},...,\msf{ID}_{i_m}$; 
 2) $\mcal{A}$ may make polynomial number of queries to a decryption oracle to obtain the corresponding plaintexts of the chosen ciphertext $C_{i}$ with $\msf{ID}_i$; 
3) $\mcal{A}$ outputs two chosen messages $(M_1$, $M_2)$ and the target identity, which is not queried for the private key, and is given a {\it challenge ciphertext} $C^{*}=\textbf{Encrypt}(\msf{param},\msf{ID}^{*},M_{b})$ on message $M_{b}$ according to a random bit $b\in\{0,1\}$; 
4) $\mcal{A}$ makes another polynomial number of queries to extract the private keys and decrypt for the plaintexts by the given identities and ciphertexts~(Restriction: The queried identites and ciphertexts should be different from $\msf{ID}^{*}$ and $C^{*}$); 
and 5) eventually, $\mcal{A}$ outputs a bit $b'\in\{0,1\}$. 
If $b=b'$, $\mcal{A}$ wins the game.}

\subsubsection{{Proof on Dual Encryptions}}\label{subsubsec:Enc_Proof}
\begin{figure}[t!] 
\begin{center}
\includegraphics[width=1\columnwidth]%
{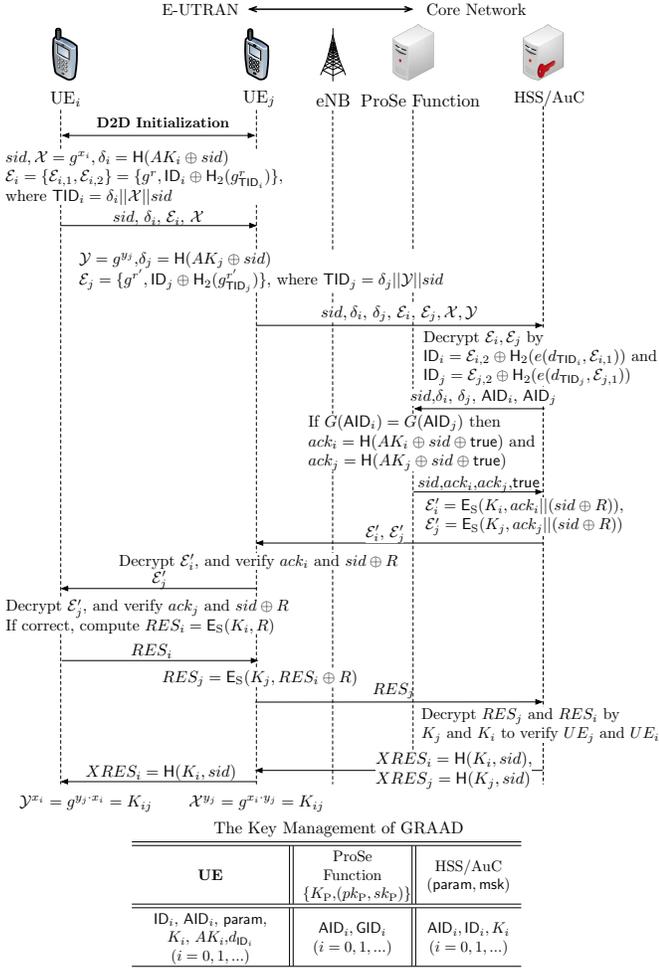} \caption{Network-covered group anonymous D2D authentication protocol with the assistance of CN, including ProSe function and HSS/AuC. Here, $s_i$, $x_i$, $y_j$, and $R$ are randomly selected number by $\UE_i$, $\UE_j$, and HSS/AuC, respectively.}
\label{fig:d2d_auth_cn}
\end{center}
\end{figure}

{To support traceability of the group anonymous AKE in network-absent D2D communications, we will adopt the concept of proof on dual encryptions. The transcripts of each authentication session is anonymous to the outsiders, including ProSe function, as group anonymity is guaranteed. We leave a trapdoor for the ProSe function to open the messages of the session by encrypting the message in a session with the public key of each participant device and the public key of ProSe function and proving the encryptions on the same messages. Before introducing the \textit{proof on dual encryptions}, we introduce two encryption algorithms, i.e., key-private and Linear public-key encryptions. 
The procedures of {\it key-private encryption} are as follows: 1) the public key and the private key of a user are given by $pk=(g,p,X)$ and $sk=(g,p,x)$, respectively, and 2) the ciphertext of the key-private encryption on a message $M$ becomes $C_1=(g^{y},M \cdot X^{y}) = (Y,C)$. One can decrypt the message by $M=C/Y^{x}$.}
%
{
The {\it Linear encryption}~\cite{GS_BBS04} is underlaid on the decision linear problem and shown as follows: 1) the public key and the private key of $T$ are given by $pk=(u,v,h)$ and $sk=(\hat{x},\hat{y})$, respectively, and 2) the ciphertext of the linear encryption on a message $M$ becomes $C_2 = (M \cdot h^{\beta_1+\beta_2}, u^{\beta_1}, v^{\beta_2})= (\hat{C},T_1,T_2)$. One can decrypt the message by $M=\hat{C}/\left(T_{1}^{\hat{x}}\cdot{}T_{2}^{\hat{y}}\right)$.}
%
{
The proof of dual encryptions, i.e., the key-private encryption and the linear encryption, and its verification can be done as the following two algorithms.
\begin{itemize}[leftmargin=*]
\item{\bf EncProof}$(C_1,C_2,\beta_1,\beta_2,y):$   
First, the prover randomly selects $r_{\beta_1}$, $r_{\beta_2}$, and $r_y$, and computes $R_1\!=u\!^{r_{\beta_1}}$, $R_2\!=\!v^{r_{\beta_2}}$, and $R_3\!=\!e(h,g)^{r_{\beta_1}+r_{\beta_2}}\cdot e(X,g)^{-r_y}$. Then it computes $c\!=\!\Hf(C_1,$ $C_2,$ $R_1,$ $R_2,$ $R_3,$ $T_1,$ $T_2,$ $X)$ and $s_{\beta_1}=r_{\beta_1}+c\beta_1$, $s_{\beta_2}=r_{\beta_2}+c\beta_2$, and $s_{y}=r_{y}+cy$. After that, the prover outputs $(c,s_{\beta_1},s_{\beta_2},s_{y},R_1,R_2,R_3)$ as the proof to the verifier. 
\item{\bf EncVer}$(C_1,C_2,c,s_{\beta_1},s_{\beta_2},s_{y},R_1,R_2,R_3):$ The verifier then checks if $C_1$ and $C_2$ are the ciphertexts on the same plaintext by $R_1=u^{s_{\beta_1}}\cdot T^{-c}_1$, $R_2=v^{s_{\beta_2}}\cdot T^{-c}_2$, $R_3=e(h,g)^{s_{\beta_1}+s_{\beta_2}}\cdot e(X,g)^{-s_{y}}\cdot e(C/\hat{C},g)^{c}$, and $c=\Hf(C_1,C_2,R_1,R_2,R_3,T_1,T_2,X)$. If the equations hold, the ciphertexts produced by the prover are the encryptions on the same plaintext, and this algorithm outputs $\msf{true}$ or $\msf{false}$ otherwise.
\end{itemize}
}
{\subsection{Key Management and User Registration}\label{subsec:key_management_user_registration}}
{The key management of the proposed GRAAD inherits the key management of the conventional security architecture of LTE, where each $\textrm{UE}_i$ shares a long-term secret key $K_i$ with HSS/AuC after registration. In addition, the key management on UEs, ProSe function, and HSS/AuC for the security of ProSe is considered as shown in Fig.~\ref{fig:d2d_auth_cn}. On HSS/AuC, the real identity $\ID_i$, the corresponding application identity $\AID_i$ and the shared secret $K_i$ of each UE are stored. Additionally, the HSS/AuC generates a master secret key, $\msf{msk}=s$, for randomly selected $s\in\mbb{Z}_{p}$ and the corresponding public parameters, $\msf{param}$, by {\bf Setup} (as introduced in Sec.~\ref{subsubsec:IBE}) to build an BF-IBE system. On the ProSe function, $\AID_i$ and the corresponding group identity $\GID_i$ of each UE are managed. Besides that, the ProSe function is associated a unique secret key $K_{\textrm{p}}$ and generates a public/private key pair of Linear encryption as $\{pk_{\textrm{P}}=(u,v,h),sk_{\textrm{P}}=(\hat{x},\hat{y})\}$. HSS/AuC issues each UE$_i$ a BF-IBE user private key $d_{\ID_i}=s\cdot{}Q_{\UID_i}$ by {\bf Extract} in Sec.~\ref{subsubsec:IBE}, where $Q_{\UID_i}=\Hf_{1}(\UID_i)$ and $\UID_i=\GID_i||\AID_i$.}


%
\subsection{Group Anonymous AKE for Network-covered D2D Communication~(CN-GD2C)} \label{subsec:CN-GD2C}

{The notations are introduced in Table~\ref{table:notations} and the key management is introduced in Sec.~\ref{subsec:key_management_user_registration} for the proposed protocol. The details of the proposed CN-GD2C protocol are as follows.}


\begin{enumerate}[leftmargin=*]
\item {After negotiating the parameters for D2D communication, $\UE_i$ randomly selects a session identity $sid\in\mbb{Z}_{p}$ and $(x_i,r)\in\mbb{Z}_{p}^{2}$. It then computes $\mcal{X}=g^{x_i}$, $\delta_i=\Hf(AK_i\oplus{}sid)$ and $\mcal{E}_i=\{g^{r},\ID_i\oplus{}\Hf_2(g_{\TID_i}^{r})\}$ and send them to $\UE_j$, where $\TID_i=\delta_i||\mcal{X}||sid$ and $\mcal{E}_i$ is the BF-IBE ciphertext.}

\item {$\UE_2$ keeps $\mcal{X}$ and computes $\mcal{Y}=g^{y_j}$, $\delta_j=\Hf(AK_j\oplus{}sid)$, and $\mcal{E}_j=\{g^{r'},\ID_j\oplus{}\Hf_2(g_{\TID_j}^{r'})\}$ with randomly selected $y_j,r'$, where $\TID_j=\delta_j||\mcal{Y}||sid$ and $\mcal{E}_j$ is the BF-IBE ciphertext. It then sends $sid,\delta_i,\delta_j,\mcal{E}_i,\mcal{E}_j,\mcal{X},\mcal{Y}$ to HSS/AuC via eNB.} 

\item {The HSS/AuC first decrypts $\mcal{E}_i$ and $\mcal{E}_j$ by $\ID_i=\mcal{E}_{i,2}\oplus{}\Hf_2(e(d_{\TID_i},\mcal{E}_{i,1}))$ and $\ID_j=\mcal{E}_{j,2}\oplus{}\Hf_2(e(d_{\TID_j},\mcal{E}_{j,1}))$, where $\TID_i=\delta_i||\mcal{X}||sid$ and $\TID_j=\delta_j||\mcal{Y}||sid$. $d_{\TID_i}$ and $d_{\TID_j}$ can be produced by the HSS/AuC with $\msf{msk}$. The HSS/AuC then finds the corresponding application identities $\AID_i$ and $\AID_j$ of $\ID_i$ and $\ID_j$ and sends $sid,\delta_i,\delta_j,\AID_i,\AID_j$ to the ProSe function to check the group information. If $G(\AID_i)=G(\AID_j)$, then the ProSe function computes $ack_i=\Hf(AK_i\oplus{}sid\oplus{}\msf{true})$ and $ack_j=\Hf(AK_j\oplus{}sid\oplus{}\msf{true})$, where $G(\AID)$ returns the belonging group of $\AID$ and $AK_i=\AID_i\oplus{}\GID_i\oplus{}K_{\textrm{P}}$. Once received $sid,ack_i,ack_j$, and $\msf{true}$, the HSS/AuC confirms $\AID_i$ and $\AID_i$ belong to the same group and sends $\mcal{E}'_i=\Ef(K_i,ack_i||(sid\oplus{}R))$ and $\mcal{E}'_j=\Ef(K_j,ack_j||(sid\oplus{}R))$ to $\UE_j$. }

\item {$\UE_j$ first decrypts $\mcal{E}'_i$ with $K_i$ to obtain $ack_i||(sid\oplus{}R)$ and verify $ack_i\stackrel{?}{=}\Hf(AK_i\oplus{}sid\oplus{}\msf{true})$. If so, $\UE_j$ extracts $R$ from $sid\oplus{}R$ and sends $\mcal{E}'_i$ to $\UE_i$.}

\item {$\UE_i$ decrypts $\mcal{E}'_j$ with $K_j$ to obtain $ack_i||(sid\oplus{}R)$ and verify $ack_j$ the same as the previous step. Then, $\UE_i$ extracts $R$ and sends $RES_i=\Ef(K_i,R)$ to $\UE_j$. Once received $RES_i$, $\UE_j$ computes $RES_j=\Ef(K_j,RES_i\oplus{}R)$ and sends to the HSS/AuC.}

\item {The HSS/AuC decrypts $RES_j$ with $K_j$ to obtain $RES_i\oplus{}R$. It then obtains $RES_i$ from $RES_i\oplus{}R$ by $R$ and decrypts $RES_i$ with $K_i$ to check if it is equal to $R$. If so, HSS/AuC sends $XRES_i=\Hf(K_i,sid)$ and $XRES_j=\Hf(K_j,sid)$ to $\UE_j$, and $\UE_j$ forwards $XRES_i$ to $\UE_i$. $\UE_i$ and $\UE_j$ accept the authenticated key exchange session for the following D2D communication between them according the verification on $XRES_i$ and $XRES_j$. Finally, $\UE_i$ and $\UE_j$ computes the same session key by $K_{ij}=\mcal{Y}^{x_i}=g^{y_j\cdot{}x_i}=\mcal{X}^{y_j}=g^{x_i\cdot{}y_j}$, respectively.}
\end{enumerate}

\subsection{Group-anonymous AKE for Network-absent D2D Communication (NA-GD2C)}\label{subsec:NA-GD2C}

This section presents a group anonymous AKE for network-absent D2D communication~(NA-GD2C) protocol with traceability, where only two devices are involved in the protocol. {Specifically, the objective of NA-GD2C protocol is to conceal the group information of both devices from outsiders and CN, except for a trusted authority that is granted to reveal the group information of users and not a part of CN. As the dispute is arisen in a session, designated authorities, i.e., ProSe function and HSS/AuC, can engage to trace the identities of the originators. Nonetheless, the identity of every UE is revocable by announcing the revoked identities in the system. The NA-GD2C protocol achieves the aforesaid goals based on the techniques of {\it $k$-anonymous secret handshakes}, {\it identity-based encryption}, and {\it non-interactive zero-knowledge proof}.}

{In the following subsections, we describe the design intuition of group anonymous protection based on $k$-anonymous secret handshakes and identity-based encryption. Then, we present the propose the NA-GD2C protocol based on the proposed group anonymous protection technique.}

\subsubsection{{Group anonymity by $k$-anonymous Secret Handshakes}}\label{subsubsec:grp_anon_k-anon_SH} The $k$-anonymous secret handshakes~(SH) can achieve adjustable {\em group anonymous authentication} where the adversary exists with the probability of $\frac{1}{k}$ to identify the group information of given user pairs~\cite{SH_TX04}. 
Moreover, the $k$-anonymous SH enjoys the property of revocability since it utilizes user certificates, which are reusable and can be revoked by announcing certificate revocation list~(CRL). {Compared to the unlinkable secret handshakes~\cite{SH_TX06} by group signatures and group key agreement, $k$-anonymous SH needs less computation costs.} SH supports each user to authenticate to the others according to the possessed group information but not identity information~\cite{SH_BDSS03,SH_CJT04,SH_TX04}. Namely, each user belonging to a group can only successfully authenticate to the other users in the same group. Otherwise, the authentication process does not leak any information to the counterpart or eavesdroppers who do not belong to the same group. {However,the communication costs of $k$-anonymous SH is linear to the anonymity degree, i.e., $k$, for exchanging the public keys of selected user pairs in the protocol. Hence, this work shows an enhanced $k$-anonymous SH by applying identity-based encryption in the design. The public keys of the selected user pairs are replaced with the identities, which can be derived by constant number of variables, of them. We propose the following four functions to achieve $k$-anonymous SH with constant communication cost in the proposed NA-GD2C.
\begin{itemize}[leftmargin=*]
\item \textbf{gSelect}$(\mbf{G},U,w,N_{u},N_{v}):$ $\mbf{G}$ is divided into $\{\mbf{G}_1$,...,$\mbf{G}_{w}\}$, where $\mbf{G}_z=\{\mbf{G}_{z_0},...,\mbf{G}_{z_{\lceil m/(w-1) \rceil}} \}$, where $U\in\mbf{G}_{i_{u}}$ for some $0\leq{}i\leq{}w-1$ and $0\leq{}u\leq{}m/(w-1)$. Set $\eta=f_{1}(N_{u},N_{v},0)$, $x=f_{1}(N_{u},N_{v},1,)$, and $y=u+r\cdot{}(m/w)$, where $r$ is randomly selected from $\{0,...,\lfloor(p+1)\cdot{}(w/m)\rfloor\}$, where $p$ is a large prime. Solve $\theta_1$ with $(y,\eta,x)$ such that $y=\eta{}\cdot{}x+\theta_1 \mod p$. For $z=0$ to $w-1$~(except $z=i$), set $y=\eta\cdot{}f_1(N_{u},N_{v},1,z)+\theta_1 \mod p$ and $s_{z}=y \mod m/w$. Then, compute $\sigma_{g}=\Hf(s_0,...,s_{w-1})$ and output $(\theta_1,\sigma_g)$.
％％％％％％％％％％％％％％％％％％％％％％％％％％％％％％
\item \textbf{uSelect}$(\bar{\textbf{G}},X,w,N_{u},N_{v}):$ $\bar{\mbf{G}}$ is divided into $\{\bar{G}_{z_{s_z}}\}_{z=0}^{w-1}$, where $\bar{G}_{z_{s_z}}\in\mbf{G}$ and $X$ is the $\lambda$-th member of $\bar{G}_{a_{s_a}}$ for some $0\leq{}a\leq{}w-1$ and $0\leq{}\lambda\leq{}|\bar{G}_{a_{s_a}}|-1$. Set $\eta=f_1(N_{u},N_{v},2)$ and $x=f_1(N_{u},N_{v},3,a,s_a)$, and $y=\lambda+r\cdot{}|\bar{G}_{a_{s_a}}|$, where $r$ is selected randomly from $\{0,...,\lfloor (p+1)/|\bar{G_{a_{s_a}}}| \rfloor\}$. Solve $\theta_2$ with $(y,\eta,x)$ such that $y=\eta\cdot{}x+\theta_2 \mod p$. For $z=0$ to $w-1$~(except $z=a$), set $y=\eta\cdot{}f_1(N_{u},N_{v},3,z,s_z)+\theta_2 \mod p$ and $\lambda_{z}=y \mod |\bar{G}_{z_{s_z}}|$. It then computes $\sigma_{u}=\Hf(\lambda_0,...,\lambda_{w-1})$ and outputs $(\theta_2,\sigma_{u})$.
\item \textbf{gSelectVer}$(\mbf{G},w,N_{u},N_{v},\theta_1,\sigma_g):$ First, for $z=0$ to $w-1$, calculate $\eta=f_1(N_{u},N_{v},0)$, $y=\eta\cdot{}f_1(N_{u},N_{v},1,z)+\theta_1 \mod p$, and $s_z=y \mod m/w$. After that, check whether $\sigma_g\stackrel{?}{=}\Hf(s_0,...,s_{w-1})$. If so, output $(s_0,...,s_{w-1})$. 
\item \textbf{uSelectVer}$(\bar{\mbf{G}},w,N_{u},N_{v},\theta_2,\sigma_u):$ Parse $\bar{\mbf{G}}$ as $\{\bar{G}_{z_{s_z}}\}_{z=0}^{w-1}$. Afterwards, for $z=0$ to $w-1$, calculate $\eta=f_1(N_{u},N_{v},2)$, $x=f_1(N_{u},N_{v},3,z,s_{z})$, $y=\eta\cdot{}x+\theta_2 \mod p$, and $\lambda_{z}=y \mod |\bar{G}_{z_{s_z}}|$. Then, check whether $\sigma_u\stackrel{?}{=}\Hf(\lambda_0,...,\lambda_{w-1})$. If so, output $\{X_{z_{s_z},\lambda_z}\}_{z=0}^{w-1}$, where $X_{s_{s_z},
lambda_z}$ denotes the $\lambda_z$-th user in $\bar{G}_{z_{s_z}}$.
\end{itemize}
These four functions are to guarantee the exchanged group and user information being selected randomly for $k$-anonymous protection in NA-GD2C. 
}
\subsubsection{Proposed NA-GD2C Protocol}

\begin{figure}[t!] 
\begin{center}
\includegraphics[width=1\columnwidth]%
{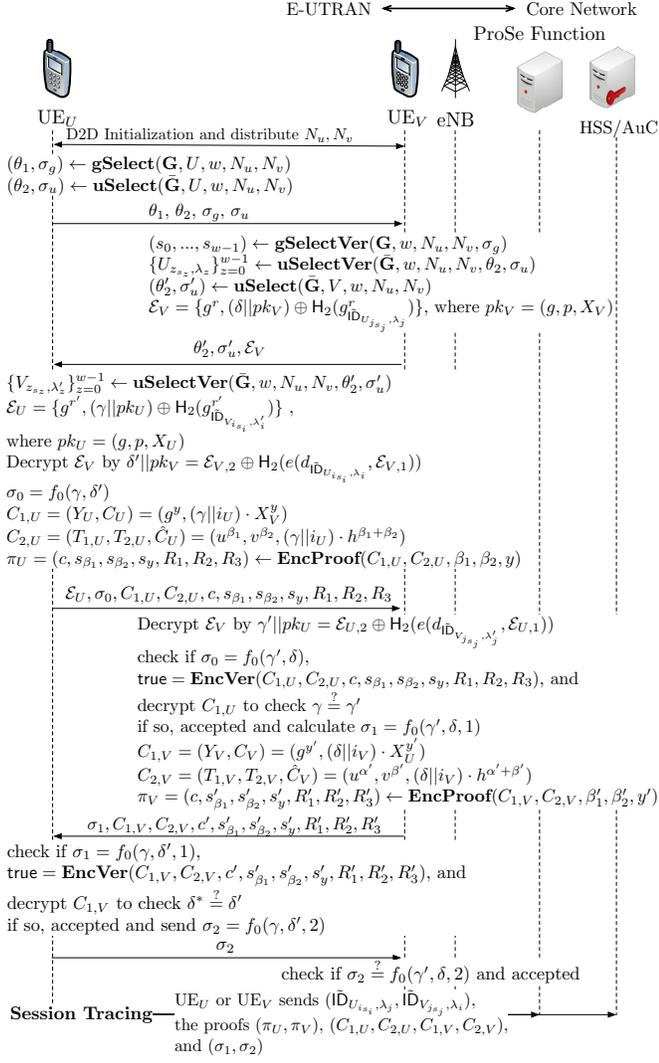} \caption{Network-absent group anonymous D2D authentication~(NA-GD2C) protocol.}
\label{fig:d2d_auth_na}
\end{center}
\end{figure}

Let us consider $n$ UEs, $\mathbf{U}=\{\UE_1,...,\UE_n\}$, that belong to $m$ different {application} groups, $\mathbf{G}=\{G_1,...,G_m\}$, in D2D communication. {For $\UE_i$ with an assigned identity $\ID_i$, the identity-based user private key $d_{\ID_i}$ is issued by HSS/AuC as introduced in Sec.~\ref{subsec:key_management_user_registration}. Additionally, $\ID_i$ generates a public/private key pair, $(pk_i,sk_i)$, of key-private public key encryption as introduced in Sec.~\ref{subsubsec:Enc_Proof}.}
Each UE belongs to one group only, and $G(U_i)=G_{j}$ means the belonging group of $U_i$ is $G_{j}$.

{We consider two UEs, an $\UE_{U}\in\mathbf{U}$ belongs to the group $G(\UE_U)=G_{i_{s_i}}$ and an $\UE_V\in\mathbf{U}$ belongs to the group $G(\UE_V)=G_{j_{s_j}}$. 
For D2D communication between $\UE_U$ and $\UE_V$, they want to authenticate each other to check whether they are legal users and belong to the same group.}
The protocol of proposed group-anonymous D2D communication is shown in Fig.~\ref{fig:d2d_auth_na} and described as follows. 

\begin{enumerate}[leftmargin=*]
\item {In the beginning, $\UE_U$ and $\UE_V$ negotiate the parameters of D2D communication and exchange two random numbers $N_{U}$ and $N_{V}$ selected by them, respectively. Afterwards, $\UE_U$ generates $\theta_1$, $\theta_2$, $\sigma_g$, and $\sigma_u$ by $(\theta_1,\sigma_g)\!\leftarrow{}\! \textbf{gSelect}(\mbf{G},\UE_U,w,N_U,N_V)$ and $(\theta_2,\sigma_u)\!\leftarrow{}\! \textbf{uSelect}(\bar{\mbf{G}},\UE_U,w,N_U,N_V)$ and sends them to $\UE_V$, where $\textbf{gSelect}$ and $\textbf{uSelect}$ are introduced in Sec.~\ref{subsubsec:grp_anon_k-anon_SH}.}

\item {$\UE_V$ generates $(s_0,...,s_{w-1})\!\leftarrow{}\!\textbf{gSelectVer}(\bar{\mbf{G}},w,$ $N_u,$ $N_v,$ $\sigma_g)$, $\{U_{z_{s_z},\lambda_z}\}_{z=0}^{w-1}\!\leftarrow{}\!\textbf{uSelectVer}(\bar{\mbf{G}},$ $V,$ $w,$ $N_{u},$ $N_{v},$ $\theta'_2,$ $\sigma'_{u})$, and encrypts a randomly selected $\delta\in\mbb{Z}_p$ and $pk_V$ as $\mcal{E}_{V}=\{g^{r},(\delta||pk_{V})\oplus{}\Hf_2(g^{r}_{\tilde{\ID}_{U_{j_{s_j},\lambda_j}}})\}$, where $r\in\mbb{Z}_p$ is selected at random. It then sends $(\theta'_2,\sigma'_{u},\mcal{E}_{V})$ to $\UE_U$.}

\item  {$\UE_U$ generates $\{V_{z_{s_z},\lambda'_z}\}_{z=0}^{w-1}\!\leftarrow{}\!\textbf{uSelectVer}(\bar{\mbf{G}}$, $w,$ $N_{u},$ $N_{v},$ $\theta'_2,$ $\sigma'_{u})$, and encrypts a randomly selected $\gamma\in\mbb{Z}_p$ and $pk_{U}$ as $\mcal{E}_{U}=\{g^{r'},(\gamma||pk_{U})\oplus{}\Hf_2(g^{r'}_{\tilde{\ID}_{V_{i_{s_i},\lambda'_{i}}}})\}$. It then decrypts $\mcal{E}_{V}$ by $\delta'||pk_{V}'=\mcal{E}_{V,2}\oplus{}\Hf_{2}(e(d_{\tilde{\ID}_{U_{i_{s_i},\lambda_i}}},\mcal{E}_{V,1}))$ to obtain $\delta'$ and $pk'_{V}$. Afterwards, $\UE_{U}$ computes $\sigma_0=f_{0}(\gamma,\delta')$ and encrypts $\gamma$ with $pk_{\textrm{P}}$ and $pk_{U},sk_{U}$ as $C_{1,U}=(Y_{U},C_{U})=(g^{y},(\gamma||i_{U})\cdot{}X_{V}^{y})$ and $C_{2,U}=(T_{1,U},T_{2,U},\hat{C}_{U})=(u^{\beta_1},v^{\beta_2},(\gamma||i_{U})\cdot{}h^{\beta_1+\beta_2})$, where $i_{U}$ is the group index of $U$ such that $U\in{}G_{i_{U}}$. It then obtains $\pi_{U}=(c,s_{\beta_1},s_{\beta_2},s_{y},R_1,R_2,R_3)$ by running $\textbf{EncProof}(C_{1,U},C_{2,U},\beta_1,\beta_2,y)$ as in Sec.~\ref{subsubsec:Enc_Proof} and sends $\mcal{E}_{U},\sigma_{U},C_{1,U},C_{2,U},c,s_{\beta_1},s_{\beta_2},s_{y},R_1,R_2,$ and $R_3$ to $\UE_V$.} 

\item {$\UE_V$ first decrypts $\mcal{E}_V$ by $\gamma'||pk_{U}=$ $\mcal{E}_{U,2}\oplus{}$ $\Hf_2(e(d_{\tilde{\ID}_{V_{j_{s_j},\lambda_j}}},$ $\mcal{E}_{U,1}))$ and checks if $\sigma_{U}\stackrel{?}{=}f_{0}(\gamma',\delta)$, $\msf{true}\stackrel{?}{=}\textbf{EncVer}(C_{1,U},C_{2,U},c,s_{\beta_1},s_{\beta_2},s_{y},R_1,R_2,R_3)$, and $\gamma'||i_{U}\stackrel{?}{=}C_{U}/Y_{U}^{x_{V}}$. If so, it accepts $\UE_U$ belonging to its group and computes $\sigma_1=f_{0}(\gamma',\delta,1)$, $C_{1,V}=(Y_{V},C_{V})=(g^{y'},(\delta||i_{V})\cdot{}X_{U}^{y'})$, $C_{2,V}=(T_{1,V},T_{2,V},\hat{C}_{V})=(u^{\alpha'},v^{\beta'},(\delta||i_{V})\cdot{}h^{\alpha'+\beta'})$, and $\pi_{V}=(c,s'_{\beta_1},s'_{\beta_2},s'_{y},R'_1,R'_2,R'_3)\!\leftarrow{}\!\textbf{EncProof}(C_{1,V},C_{2,V},\beta'_1,\beta'_2,y')$, where $i_V$ is the group index of $V$ such that $V\in{}G_{i_V}$. After that, $\UE_V$ sends $\sigma_1,C_{1,V},C_{2,V},c',s'_{\beta_1},s'_{\beta_2},s'_{y},R'_1,R'_2,$ and $R'_3$ to $\UE_U$.}

\item {$\UE_U$ checks if $\sigma_1\stackrel{?}{=}f_{0}(\gamma,\delta',1)$, $\msf{true}\stackrel{?}{=}\textbf{EncVer}(C_{1,V},C_{2,V},c',s'_{\beta_1},s'_{\beta_2},s'_{y},R'_1,R'_2,R'_3)$, and $\delta'||i_{V}\stackrel{?}{=}C_{V}/Y_{V}^{x_{U}}$. If so, it accepts $\UE_V$ belonging to its group and computes $\sigma_2=f_{0}(\gamma,\delta',2)$. $\UE_{U}$ sends $\sigma_2$ to $\UE_V$. $\UE_V$ then checks the correctness of $\sigma_2$ and stores it. Finally, $\UE_U$ and $\UE_V$ share the same session key $K_{U,V}=f_{0}(\gamma,\delta,3)$.}
\end{enumerate}

\subsection{NA-GD2C with Traceability} \label{subsubsec:Traceability_GD2C}

{In case of dispute or management requirement, $\UE_U$ or $\UE_V$ sends $(\tilde{\ID}_{U_{i_{s_i},\lambda_i}},\tilde{\ID}_{V_{j_{s_j},\lambda_j}})$, $(\pi_{U},\pi_{V})$, and $\sigma_1,\sigma_2$ to the ProSe function to prove the participation of $\UE_{U}$ and $\UE_{V}$ in the specified D2D communication session as follows. The ProSe function first checks the correctness of $(C_{1,U},C_{2,U},C_{1,V},C_{2,V})$ by $\textbf{EncVer}$ with $(\pi_{U},\pi_{V})$ as in Sec.~\ref{subsubsec:Enc_Proof}. It then decrypts $C_{2,U}$ and $C_{2,V}$ to obtain $\gamma$ and $\delta$ by $\hat{C}_{U}/(T_{1,U}^{\hat{x}}\cdot{}T_{2,U}^{\hat{y}})$ and $\hat{C}_{V}/(T_{1,V}^{\hat{x}}\cdot{}T_{2,V}^{\hat{y}})$ to verify $\sigma_0$, $\sigma_1$ or $\sigma_2$.}

\subsection{{Membership Management}}
{The membership of the proposed two protocols can be managed as follows. For CN-GD2C, once any member of an application group joins or leaves, the HSS/AuC and {ProSe function} will update the membership information. Since every authentication needs to interact with HSS/AuC {and ProSe function}, each membership update will immediately take effect in the following D2D authentication sessions. For NA-GD2C, every authentication uses the {identities} of UEs in different application groups. Hence, HSS/AuC will issue {the corresponding user private key} for each new UE, and revoke {the identities} of revoked UEs and list them on the revocation list in public. In summary, the costs of membership maintenance are both constants for CN-GD2C and NA-GD2C regardless of the number of groups and UEs in the system.}

\section{Security Analysis}\label{sec:security_analysis}
In this section, we prove the security of the proposed protocols based on the security definition in Sec.~\ref{subsec:sec_model_def}.

\subsection{Security Analysis of CN-GD2C Protocol}
In the network-covered D2D communications, traceability and revocability are readily guaranteed as all communications of UEs are performed under the control of CN. 
Hence, we prove that the proposed CN-GD2C protocol achieves
the mutual authentication, secure key exchange for end-to-end security, and identity and group anonymity as follows.

\begin{theorem}{(Mutual Authentication)}\label{theorem:mutual_auth_cn}
{The proposed CN-GD2C protocol is a {\em mutual authentication} protocol between two UEs if $\Ef$ is a pesudorandom permutation and $\Hf$ is a pseudorandom function.}
\end{theorem}

\begin{IEEEproof} 
{The CN-GD2C achieves mutual authentication between $\UE_i$ and $\UE_j$ by the verification of $(RES_i,RES_j)$ by HSS/AuC and that of $(\mcal{E}'_i,\mcal{E}'_j)$ and $(XRES_i,XRES_j)$ by $\UE_i$ and $\UE_j$, respectively. We define $\mbf{Adv}_{\mcal{A}}^{auth}=\Pr[\mbf{E}_1\vee{}\mbf{E}_2\vee\mbf{E}_3\vee\mbf{E}_4\vee\mbf{E}_5\vee\mbf{E}_6]$, where $\mbf{E}_1$ and $\mbf{E}_2$ are the events that $\mcal{A}$ successfully impersonates the HSS/AuC by sending a verifiable $\mcal{E}'_i$ and $\mcal{E}'_j$, $\mbf{E}_3$ is the event that $\mcal{A}$ successfully impersonates $\UE_i$ by sending a correct $RES_i$, $\mbf{E}_4$ is the event that $\mcal{A}$ impersonates $\UE_j$ by sending a correct $RES_j$, $\mbf{E}_5$ and $\mbf{E}_6$ are the events that $\mcal{A}$ impersonates HSS/AuC by sending $XRES_i$ or $XRES_j$, accepted by $\UE_i$ or $\UE_j$, respectively. Here, $\Pr[\mbf{E}_1\vee{}\mbf{E}_2\vee\mbf{E}_3\vee\mbf{E}_4\vee\mbf{E}_5\vee\mbf{E}_6] \leq \Pr[\mbf{E}_1] + \Pr[\mbf{E}_2] + \Pr[\mbf{E}_3] + \Pr[\mbf{E}_4] + \Pr[\mbf{E}_5] + \Pr[\mbf{E}_6]$. Hence, we prove that the probabilities are negligible to guarantee that $\mbf{Adv}_{\mcal{A}}^{auth}$ is negligible for mutual authentication security as follows.}

{ For the case of $\Pr[\mbf{E}_1]$, we consider $\Ef$ as a pseudorandom permutation and the advantage $\epsilon$ of distinguishing a pseudorandom permutation and random permutation is negligible. A simulator $\mcal{S}$ is given a function ${\Ef}_{b}$ by the challenger according to a random bit $b\in\{0,1\}$, where ${\Ef}_{0}$ is a pseudorandom permutation and ${\Ef}_{1}$ is a random permutation. $\mcal{S}$ simulates the protocol and interacts with $\mcal{A}$ by the given ${\Ef}_{b}$. We say that $\mcal{A}$, acting as a HSS/AuC, can send out a correct $\mcal{E}'_i$ to be accepted by $\mcal{S}$ with the probability $\epsilon'$. Since the probability that $\mcal{A}$ sends a correct $\mcal{E}'_i$ with a random permutation, i.e., ${\Ef}_{1}$, is negligible, $\mcal{S}$ outputs a guess $b'=1$ when $\mcal{E}'_i$ sent by $\mcal{A}$ is accepted. Otherwise, $\mcal{A}$ outputs $b'=0$ or $b'=1$ randomly. From the above, we have that $\Pr[b=b']=$ $\Pr[b=b',b=0] + $ $\Pr[b=b',b=1]=$ $\Pr[b=b'|b=0]\Pr[b=0]$ + $\Pr[b=b'|b=1]\Pr[b=1] = (\frac{1}{2}+\epsilon')\frac{1}{2} + \frac{1}{2} \times \frac{1}{2}=$ $\frac{(1+\epsilon')}{2}$. If $b=b'$, ${\Ef}_{b}$ is distinguished. Obviously, $\epsilon \geq \Pr[b=b']-\frac{1}{2}=\frac{\epsilon'}{2}$. Hence, $\epsilon'$ is negligible since $\epsilon$ is also negligible based on pseudorandom permutation. We can prove $\Pr[\mbf{E}_2]$, $\Pr[\mbf{E}_3]$, $\Pr[\mbf{E}_4]$, $\Pr[\mbf{E}_5]$, and $\Pr[\mbf{E}_6]$ are also negligible by the similar way. Based on the above proof, we conclude that $\mbf{Adv}_{\mcal{A}}^{auth}=\Pr[\mbf{E}_1\vee{}\mbf{E}_2\vee\mbf{E}_3\vee\mbf{E}_4\vee\mbf{E}_5\vee\mbf{E}_6]$ is negligible and mutual authentication between $\UE_1$ and $\UE_2$ holds.}
\end{IEEEproof}

\begin{theorem}{(Secure Key Exchange)}\label{theorem:sec_key_exchange_cn}
{The proposed CN-GD2C protocol is an {\em authenticated key exchange} protocol between two UEs if Diffie-Hellman assumption holds.}
\end{theorem}
\begin{IEEEproof} This proof is separated as two parts, session key security and known key security. Regarding the session key security of {$K_{ij}$}, it is based on the Decisional Diffie-Hellman~(DDH) assumption, where a truly DDH tuple $(\mcal{X}=g^x,\mcal{Y}=g^y,K_{ij}=g^{xy})$ is indistinguishable from a random tuple $(g^x, g^y, g^z)$. That is, the security of DDH assumption is reducible to the security of $K_{ij}$ as follows. One can construct a simulator $\mcal{S}$ to simulate CN-GD2C by a given tuple $(g^x,g^y,g^{z})$, which is either a DDH or random tuple. If an adversary $\mcal{A}$, interacting with $\mcal{S}$, can distinguished the generated session key from a random string in the simulated protocol with an advantage of $\epsilon$. $\mcal{S}$ is able to distinguish the given DDH tuple with an advantage of $\epsilon'$. The probability of distinguishing the session key or a random string is $1/2+\epsilon$. The probability of distinguishing the DDH tuple from a random tuple by the guessing result from $\mcal{A}$ is $(1+\epsilon)/2$. Obviously, $\epsilon'$ is greater than or equal to $(1+\epsilon)/2$ and $\epsilon'$ is negligible~\cite{DDH_Boneh}. Hence, $\epsilon$ is also negligible.

In terms of known key security, we have to ensure that the DDH tuple $(\mcal{X}=g^x,\mcal{Y}=g^y)$ actually contributes to produce a unique session key $g^{xy}$ without any material injected~\cite{SKSecurity_LMQSV03}. Assume that there is no collusion with the CN and any adversary. {$\mcal{X}$ and $\mcal{Y}$ are authenticated as shown in Theorem~\ref{theorem:mutual_auth_cn}}. Therefore, no adversary is able to change $\mcal{X}$ and $\mcal{Y}$ based on the security of mutual authentication.
\end{IEEEproof}

In the network-covered D2D communication, the anonymity between UE and CN is infeasible as all UEs are under the coverage of CN. 
However, in this case, it is important to guarantee the group anonymity between UEs, especially when they are in different groups. 
In the following theorem, we prove the partial group anonymity, i.e., that between UEs, of the CN-GD2C protocol. 

\begin{theorem}{({Identity and} Group Anonymity)}\label{theorem:id_grp_anony_cn}
The proposed CN-GD2C protocol is a {\em{identity and} group anonymous} protocol between $\UE_i$ and $\UE_j$ {if mutual authentication is guaranteed and BF-IBE is IND-CCA secure.}
\end{theorem}

\begin{IEEEproof}
{Before proving group anonymity, we should prove identity anonymity of the CN-GD2C as the sufficient and necessary condition. The CN-GD2C guarantees identity anonymity based on the IND-CCA security of BF-IBF. Thus, we have to prove the identity anonymity of the CN-GD2C is reducible to the key-private security of BF-IBF. One can create $\mcal{S}$, who is an attacker of the IND-CCA security of BF-IBE, simulates the CN-GD2C interacting with an identity anonymous attacker $\mcal{A}$. A challenger $\mcal{C}_{BF-IBE}$ of BF-IBE provides encryption and decryption oracles for the security game as defined in Definition~\ref{def:id_anon}. $\mcal{S}$ can simulate CN-GD2C by generating all required secret keys, except the BF-IBE user private keys of $\Pi_{i,j}^{t_1}$ and $\Pi_{j,i}^{t_2}$.} 

{We now prove the identity anonymity of $\UE_i$ and $\UE_j$ by simulating CN-GD2C with $\mcal{C}_{BF-IBE}$ and $\mcal{A}$, respectively. First, we prove the identity anonymity of $\UE_i$ as follows. $\mcal{S}$ selects two messages $\ID_{i,0}=\ID_i$ and a random string $\ID_{i,1}=\Re$ and send to $\mcal{C}_{BF-IBE}$. $\mcal{C}_{BF-IBE}$ then encrypts $\ID_{i,b}$ as $\mcal{E}_{i}$ according to $b\in\{0,1\}$. $\mcal{S}$ then simulates an AKE session of CN-GD2C with $\mcal{E}_i$. After the simulation, $\mcal{A}$ queries $\textbf{TestID}(\Pi_{i,j}^{s})$ to $\mcal{S}$ and $\mcal{S}$ returns $\ID'_{i,b}$ according to a random bit $c\in\{0,1\}$, where $\ID'_{i,0}=\ID_i$ and $\ID'_{i,1}=\Re'$, which is a random string. $\mcal{A}$ outputs $c'$ as the guess of $c$. If $c'=c$, then $\mcal{S}$ outputs $b=0$ or $b=1$ otherwise. The probability of breaking the IND-CCA security of BF-IBE is $\Pr[b=b']$ and can be rewritten as $\Pr[b=b']=\Pr[b=b'|b=0]\Pr[b=0] + \Pr[b=b'|b=1]\Pr[b=1]$. Under a real experiment~$(b=0)$, $\mcal{A}$ can win the game by successfully guessing  $(c=c')$ with probability $\frac{1}{2}+\epsilon'$, where $\mathbf{Adv}_{\mcal{A}}^{\msf{anon\_id}}=\epsilon'$. $\mcal{A}$ can only randomly guess $(c=c')$ with probability $1/2$ under a random experiment, i.e., $\Pr[b=b'|b=0]=\frac{1}{2}+\epsilon'$ and $\Pr[b=b'|b=1]=1/2$. Hence, $\Pr[b=b']=(\frac{1}{2}+\epsilon')\frac{1}{2} + \frac{1}{2}\times\frac{1}{2}=(1+\epsilon')/2$. Obviously, the probability of breaking IND-CCA security of BF-IBE is $\epsilon\geq{}\Pr[b=b']=\frac{1+\epsilon'}{2}-\frac{1}{2}=\frac{\epsilon'}{2}$. $\textbf{Adv}_{\mcal{A}}^{\msf{anon\_id}}=\epsilon'$ is negligible due to $\epsilon$ is negligible. The identity anonymity of $\UE_j$ can also be proven in the similar way as the above. Hence, the CN-GD2C is an identity anonymous AKE protocol for $\UE_i$ and $\UE_j$.}

{The group anonymity is guaranteed since the group information of UEs is only known by the ProSe function. Besides that, the ProSe function will only check the group information for the valid UEs, who can send out $ack_i$ and $ack_j$, in each AKE session. The security of $ack_i$ and $ack_j$ depends on the authorization keys, $AK_i$ and $AK_j$, issued for $\UE_j$ and $\UE_j$, and the correctness of $\AID_i$ and $\AID_j$ depends on the trust of HSS/AuC to ProSe function. We can prove that $ack_i$ and $ack_j$ can only be sent out by the legal $\UE_i$ and $\UE_j$ based on the pseudorandom function assumption in the similar concept of the proof of Theorem~\ref{theorem:mutual_auth_cn}. Hence, the identity anonymity is guaranteed to outsider attackers and the group anonymity to outsider attackers and the HSS/AuC.}
\end{IEEEproof}

\subsection{Security Analysis of NA-GD2C Protocol}
In this section, we prove that the proposed NA-GD2C protocol achieves
the mutual authentication, secure key exchange (for end-to-end security), group anonymity, and traceability and revocability.

\begin{theorem}{(Mutual Authentication)}\label{theorem:mutual_auth_na}
The proposed GD2C protocol is a mutual authentication protocol between two UEs if BF-IBE is IND-CCA secure.
\end{theorem}
\begin{IEEEproof}
{In NA-GD2C, $\UE_U$ and $\UE_V$ verify $\sigma_0$ and $\sigma_1$ for mutual authentication. The functions, $\textbf{gSelect}$, $\textbf{gSelectVer}$, $\textbf{uSelect}$, and $\textbf{uSelectVer}$ guarantee that the obtained user indices of $\UE_U$ and $\UE_V$ will include the identity of each other. Hence, if $\UE_{U}$ and $\UE_V$ are in the same group, they will select the identity of each other to encrypt $\delta$ and $\gamma$ randomly selected by $\UE_V$ and $\UE_U$, respectively. $\mcal{S}$, who is an attacker of IND-CCA secure BF-IBE, simulates CN-GD2C interacting with the attacker $\mcal{A}$ against mutual authentication. First, $\mcal{S}$ simulates $\Pi_{U,V}^{s}$ as $\UE_{U}$ interacting with $\mcal{A}$ impersonating $\UE_V$ by the BF-IBE ciphertext $\mcal{E}_{U}$ on one of two submitted $\gamma$ and $\gamma^{*}$ to the challenger $\mcal{C}$ of IND-CCA secure BF-IBE. $\mcal{E}_{U}$ is produced according to a random bit $b\in\{0,1\}$. If $b=0$, $\mcal{E}_U$ is the ciphertext of $\gamma$ or $\gamma^{*}$ otherwise. In the real experiment~(i.e., $b=0$), $\mcal{A}$ can successfully send out a correct $\sigma_1=f_{0}(\gamma,\delta')$ with the probability of $\epsilon'$. In the random experiment~(i.e, $b=1$), $\mcal{A}$ is with probability of $2^{-k}$ to produce $\sigma_1$ since $\gamma$ is never exposed to $\mcal{A}$. The probability of break IND-CCA security of BF-IBE is $\epsilon\geq{}\Pr[b=b']=\Pr[b=b'|b=0]\Pr[b=0]+\Pr[b=b'|b=1]\Pr[b=1]=(\frac{1}{2}+\epsilon')\frac{1}{2}+(\frac{1}{2}+\frac{1}{2^{k}}){}\frac{1}{2}$. Since $\epsilon$ is negligible based on IND-CCA security, $\epsilon'=\textbf{Adv}_{\mcal{A}}^{auth}$ is also negligible. We can also prove the impersonation of $\UE_{V}$ by $\mcal{A}$ in the same notion as the above. Conclusively, $\mcal{A}$ has only negligible probability to break mutual authentication in NA-GD2C based on IND-CCA secure BF-IBE.}  
\end{IEEEproof}

\begin{theorem}{(Secure Key Exchange)}
The proposed NA-GD2C protocol is a secure key exchange protocol between $\UE_{U}$ and $\UE_{V}$ if BF-IBE is IND-CCA secure.
\end{theorem}
\begin{IEEEproof}
In the NA-GD2C protocol, the exchanged session key is $K_{U,V}=f_{0}(\gamma,\delta,3)$, where $\gamma$ and $\delta$ are authentic, and $f_{0}$ is a secure hash function and can be regarded as a pseudorandom function. Hence, in the case of unknown $\gamma$ and $\delta$, the session key is indistinguishable from truly random string of equal length. The security of pseudorandom function is reducible to the security of NA-GD2C protocol. Namely, one can build an algorithm to exploit the ability of $\mcal{A}$ in the proposed NA-GD2C protocol to distinguish $\hat{\Hf}$ from a truly random function. If $\mcal{A}$ can successfully distinguish the session key from a random string, one can also successfully distinguish $f_{0}$ from a truly random function. It means that the advantage of distinguishing $f_{0}$ from a truly random function is greater than or equal to the advantage of distinguishing the session key from a random string in the proposed NA-GD2C protocol. Here, the advantage of distinguishing $f_{0}$ from a truly random function is negligible, and consequently, that of distinguishing the session key from a random string is also negligible. 
\end{IEEEproof}

\begin{theorem}{(Identity and Group Anonymity)}
The proposed NA-GD2C protocol is a $k$-anonymous authentication protocol if BF-IBE is IND-CCA secure.
\end{theorem}
\begin{IEEEproof}[Proof Sketch] 
{The identity and group information of two UEs are guaranteed based on the IND-CCA security of BF-IBE as the same concept in Theorem~\ref{theorem:id_grp_anony_cn}. Hence, the NA-GD2C achieves $k$ identity and group anonymity by perfectly hiding them in one of the $k$ selected groups and identities.}
\end{IEEEproof}
\begin{table*}[t!]\centering
\caption{Computation Costs of the Proposed Schemes}
\label{table:computation_cost_GD2C}
\begin{tabular}{|M{3.6cm}||M{3.8cm}||M{3.8cm}||M{4.5cm}|} \hline
 & CN-GD2C& NA-GD2C & NA-GD2C~(Traceability)\\\hline
Costs & $T_{IBE}+2T_{\Hf}+2T_{\Ef}+T_{DH}$ & $T_{IBE}+3T_{\Hf}$ & $T_{KPE}+T_{KPD}+T_{LIN}+8T_{EXP}+3T_{P}+6T_{\Hf}+(2w)T_{mul}+(2w+2)T_{\Hf}$\\ \hline
Computation time ($w=10$)   & 1.5716 ms & 0.824 ms  & 5.978 ms\\ \hline
Computation time ($w=50$)  & 1.5716 ms & 0.824 ms  & 6.826 ms\\ \hline
Message Length         & 852-bit~($\UE_i$) + 1832-bit~($\UE_j$)     & 1066-bit~($\UE_{U}$)+682-bit~($\UE_{V}$)      & 3574-bit~($\UE_{U}$)+ 3190-bit~($\UE_{V}$) \\\hline
\multicolumn{4}{|l|}{\pbox{16.5cm}{\vspace*{0.1cm}$T_{IBE}$: the computation time of BF-IBE encryption/decryption~(with plaintext of 128-bit and group element of 170-bit)\\
                         $T_{DH}$: the computation time of Diffie-Hellman~(with the prime of 1024-bit and the exponential of 160-bit)\\
	                     $T_{\Ef}$: the computation time of symmetric encryption~(AES)(with an input of 128-bit)\\
                         $T_{\Hf}$: the computation time of one-way hash function~(SHA-256)(with an input of 128-bit)\\
                         $T_{KPE}$: the computation time of key-private encryption~(with plaintext of 170-bit and public key of $3\times{}$170-bits)\\
                         $T_{KPD}$: the computation time of key-private decryption~(with ciphertext of $2\times$170-bit)\\
                         $T_{LIN}$: the computaion time of Linear encryption/decryption~(with plaintext of 170-bit, public key of $3\times$170-bit, and ciphertext of $3\times$170-bit)\\
                         $T_{EXP}$: the computation time of exponential operation in $\mbb{G}$\\
                         $T_{P}$: the computation time of pairing\\
	                     $T_{DH}$=0.74 ms, $T_{\Ef}=6.8\times 10^{-3} ms$, $T_{\Hf}$=0.006 ms, $T_{EXP}$=0.37 ms,
	                     $T_{IBE}$=0.806 ms, $T_{KPE}$=$T_{KPD}$=2$T_{EXP}$= 0.74 ms, $T_{LIN}$=3$T_{EXP}$=1.11 ms, $T_{P}$=0.06 ms, $T_{mul}$=0.004 ms \vspace*{0.1cm}}} \\ \hline
\end{tabular}
\end{table*}

\begin{theorem}{(Full Traceability)}
The proposed NA-GD2C protocol is with {full} traceability\footnote{The notion of full traceability introduced in~\cite{GS_BMW03} is to model a coalition of members breaking the security of traceability. That is, any of the members may break traceability by authenticating the other member without being traced to any identity of the coalition group.}, {i.e., no PPT adversary or no colluding set of UEs can successfully authenticate any UE with identities that do not exist or belong to other UEs, if the linear encrypt and key private encryption are semantically secure.}
\end{theorem}
\begin{IEEEproof} 
The security of full traceability of the NA-GD2C protocol is based on the security of the zero-knowledge proof on the same plaintext of the linear encryption and the key-private encryption.
%
The concept of proving two encryptions on a message without revealing any knowledge of the message is originally applied for building encryption schemes of stronger security~(e.g., IND-CPA$\rightarrow$IND-CCA1 and IND-CCA1$\rightarrow$IND-CCA2~\cite{Enc_NY90,Enc_Sahai99}). 
Using a non-interactive zero-knowledge~(NIZK) to prove that two encryptions are for the same message, one can simulate the decryption oracle by one of the two encryptions. The security of traceability has to consider two kinds of adversaries; 1) the one who may produce a false proof on two encryptions of different messages and 2) the one who may produce a new proof different from the original proof without the random secrets $\alpha$, $\beta$, and $x$.
 The security of traceability of the proposed GD2C consists of two security properties as follows.\\
\noindent {1) Zero knowledge:} One can prove that the prover in possession of the trapdoor to produce a false proof to prove $C$ and $\hat{C}$ are the encryptions of the same message even if they are the encryptions of different messages as follows. The prover computes $R_3=e(h,g)^{r_{\alpha}+r_{\beta}}\cdot e(Y,g)^{-r_{x}}\cdot e(\frac{M_2}{M_1},g)^{c}$ and outputs $c$ as the hash value while the input of the controlled random oracle $H$ is $(C_1,C_2,R_1,R_2,R_3,T_1,T_2,Y)$, where $M_1$ and $M_2$ are the messages of $C_1$ and $C_2$.  Hence, $c$ is considered as a random number generated by a secure hash function. The security of preventing any adversary from generating a proof on two encryptions of different messages. \\ 
\noindent {2) Simulation sound:} A challenger $\mcal{C}$ simulates the scheme of traceability and interacts with an attacker $\mcal{A}$. It is infeasible for $\mcal{A}$ to produce a proof $\pi$ to produce another proof $\pi'=(c',s'_{\alpha},s'_{\beta},s'_{x})$. Otherwise, $\mcal{C}$ is able to computes $\Delta{}c=c-c'$, $\Delta{}s_{\alpha}=s_{\alpha}-s'_{\alpha}$, $\Delta{}s_{\beta}=s_{\beta}-s'_{\beta}$, and $\Delta{}s_{x}=s_{x}-s'_{x}$, and computes $\alpha=\Delta{}s_{\alpha}/\Delta{}c$, $\beta=\Delta{}s_{\beta}/\Delta{}c$, and $x=\Delta{}s_{x}/\Delta{}c$. Then $\mcal{C}$ can extract $M$ by $\hat{C}/h^{\alpha+\beta}$ to break the Linear encryption and by $C/Y^x$ to break the key-private encryption. Assume that the probability of producing a false proof is $\epsilon$, the probability of breaking the Linear encryption is $\epsilon_{LIN}$, and the probability of breaking the key-private encryption is $\epsilon_{KP}$. From the above, we have that $\epsilon\leq \epsilon_{LIN}\times \epsilon_{KP}$. $\epsilon_{LIN}$ and $\epsilon_{KP}$ are negligible. To sum up, the probability of producing a proof on the $C_1$ and $C_2$ by $\mcal{A}$ is negligible.
\end{IEEEproof}

\section{Comparisons and Performance Evaluation}\label{sec:performance_evaluation}
In this section, {we compare the security properties of this work with two related works~\cite{TVT_D2D_ZCHQ16,TIFS_D2D_ZWYL16},} and analyze and evaluate the computation cost and authentication success rate for the proposed GD2C protocols.
{
\subsection{Comparison on Security Properties}
We compare the security properties of the proposed protocols with SeDS~\cite{TVT_D2D_ZCHQ16} and a light-weight D2D-assist data transmission protocol~(LSD)~\cite{TIFS_D2D_ZWYL16} as shown in Table~\ref{table:compar_sec_properties}. For mutual authentication, in SeDS, the message sent by eNB in step 5 can be replayed without being checked out. For end-to-end security, LDS only claims to be achieved by DH key agreement and does not concrete in the protocol. For network-absent secure D2D communication, only the proposed NA-GD2C provides. Besides that, only the proposed protocol can achieve both identity and group anonymity.

\begin{table}[t!]
\caption{Comparisons on Security Properties}
\label{table:compar_sec_properties}
\begin{center}
{\footnotesize 
\begin{tabular}{M{3cm}|c|c|c} \hline\hline
Properties & SeDS~\cite{TVT_D2D_ZCHQ16} & LSD~\cite{TIFS_D2D_ZWYL16}& Ours\\\hline\hline
\vspace{-.2mm}Mutual Authentication&\vspace{-.2mm}- &\vspace{-.2mm}\checkmark&\vspace{-.2mm}\checkmark\\\hline
\vspace{-.2mm}End-to-End Security & \vspace{-.2mm}\checkmark & \vspace{-.2mm}? &\vspace{-.2mm}\checkmark \\\hline
\vspace{-.2mm}Pseudonymity &\vspace{-.2mm}\checkmark& \vspace{-.2mm}\checkmark &\vspace{-.2mm}\checkmark \\\hline
\vspace{-.2mm}Anonymity & \vspace{-.2mm}x & \vspace{-.2mm}\checkmark& \vspace{-.2mm}\checkmark \\\hline
\vspace{-.2mm}Forward Secrecy & \vspace{-.2mm}\checkmark & \vspace{-.2mm}? & \vspace{-.2mm}\checkmark \\\hline
\vspace{-.2mm}Network-absent D2D &\vspace{-.2mm}x&\vspace{-.2mm}? & \vspace{-.2mm}\checkmark \\\hline
\vspace{-.2mm}Group Anonymity &\vspace{-.2mm}x&\vspace{-.2mm}x& \vspace{-.2mm}\checkmark\\\hline\hline
\end{tabular}
}
\end{center}
\end{table}
}
\subsection{Computation and Communication Costs}
In this section, we evaluate the computation/communication costs of the proposed schemes empirically on a smartphone of HTC One X as a testbed. The smartphone runs Android 4.1.1 mobile operating system and is equipped with 1.5 GHz quad-core ARM Cortex-A9 CPU and 1GB RAM.  The cryptographic libraries for the implementation are java pairing based cryptography (JPBC)~\cite{ISCC_DecIov11} and Java Cryptography Extension~(JCE)~\cite{JCA}. Table~\ref{table:computation_cost_GD2C} shows the total computation cost~(time) and the message length of the proposed two schemes, and the definitions of related computation times. {Regarding the message length, we build the pairing mapping by MNT curves~\cite{PB_MNT01} for 80 bits security, where the length of an element from $\mbb{G}_1$ is 170 bits and from $\mbb{G}_T$ is 340 bits. For storage cost, it only takes 682-bit in total on every $\UE_i$, where ($128\times4$) bits for ($\ID_i,\AID_i,K_i,AK_i$) and 170 bits for $d_{\ID_i}$.}  

\subsection{Authentication Success Rate of D2D Communications}
In this section, we analyze the authentication success rate (ASR) of the proposed protocols to evaluate their feasibility.  The measurement of ASR considers the effects of the arrival rate of D2D authentication requests and the residence time of a host device in eNB and that in the coverage of D2D communications~(i.e., the time that both devices are in the D2D communication coverage of each other) affect the ASR. For convenience, we name a device that initiates D2D communication as a {\it host device} and a device that is the counterpart of the host device, as a {\it target device}.

In the authentication process, 
the host device will reserve its resource for authentication (e.g., CPU) for each incoming device in first-come-first-serve~(FCFS) manner. The authentication {\em fails} 
whenever the target device departs from the coverage of D2D communication or one of the host and target devices departs from the coverage of eNB in network-covered case, before finishing authentication of the target device at the host device.

We denote the residence time of a host device in the D2D communication coverage of a target device as $t_{rd}$,
and the residence time of the host device in eNB\footnote{In D2D communication, each of two devices connect to the same eNB for authentication is not needed as long as the eNBs connect to the same CN such that the authentication messages between devices and CN can be transmitted through the eNBs that the devices attach.} as $t_{r}$.
The system authentication time is denoted by $t_a$, defined as $t_a = t_{Q} + t_{s}$,
where $t_{Q}$ is the waiting time in queue for authentication processing and $t_{s}$ is the authentication processing time.
To evaluate the ASR, we give the following assumptions.
\begin{enumerate}[leftmargin=*]
\item Every device authenticates only one device at a time. 
\item The host device residence time $t_{rd}$ in the coverage of a target device~(i.e., the residence time in D2D communication) is exponentially distributed with mean $1/\lambda_{rd}$. 
\item The host device residence time $t_r$ in the coverage of an eNB is exponentially distributed with mean $1/\lambda_{r}$. 
\item The authentication {processing} time $t_{s}$ is {constant} as $\hat{T}_{s}$. 
\item The arrival rate of host devices entering the coverage of a given target device follows Poisson random process with mean $\lambda_{t}$.
\end{enumerate}
Note that in \cite{AuthPerformance_LC03,AuthPerformance_HL09}, the expected session key life time is estimated by observing the probability that the life time of session key is greater than or equal to the residence time in new AP, where the residence time in AP is assumed to be exponentially distributed. Hence, we reasonably assume that the residence time in D2D communication and that in eNB are exponentially distributed.
%
We now analyze the ASRs of both proposed protocols for network-covered and network-absent cases.
Note that $f_\mathsf{X}(x)$ denotes the pdf of random variable $\mathsf{X}$ in the following analysis. 

\vspace{-0.5mm}
\subsubsection{ASR of NA-GD2C Protocol}
The ASR in network-absent D2D communication is mainly affected by $t_{rd}$ and $t_{a}$, and it is derived as follows.

\begin{lemma}\label{lemma:asr_NA-GD2C}
The authentication success rate of the proposed NA-GD2C protocol is 
\begin{align}\label{eq:ASR}	
	\ASR = 
		\frac{e^{-2\lambda_{rd}\hat{T}_s} (1-\lambda_{t}\hat{T}_{s})\lambda_{rd}}{\lambda_{rd}-\lambda_{t}(1-e^{-\lambda_{rd}\hat{T}_s})}. 
\end{align}

\end{lemma}

\begin{IEEEproof}
The authentication is successful when the residence time in D2D communication is greater than or equal to the system authentication, which includes the waiting time in queue and the authentication time. Hence, the ASR can be presented as 
\begin{equation}\label{eq:alpha-1}
\begin{split}
\ASR = &\Pr[t_a \leq t_{rd}] = \Pr[t_s \leq t_{rd}-t_Q] \\
= &\Pr[t_a \leq t_{rd} | t_{rd} \geq \hat{T}_s] \Pr[t_{rd} \geq \hat{T}_s]\\
&+ \Pr[t_a \leq t_{rd} | t_{rd} < \hat{T}_s]  \Pr[t_{rd} < \hat{T}_s].
\end{split}
\end{equation} 
In \eqref{eq:alpha-1}, when $t_{rd} \leq \hat{T}_s$, the probability that $t_a \leq t_{rd}$ is zero since $t_a = t_Q + \hat{T}_s$. Hence, the ASR is represented by 
\begin{align}\label{eq:ASR-1}	
	\ASR = \Pr[t_a \leq t_{rd} | t_{rd} \geq \hat{T}_s] \Pr[t_{rd} \geq \hat{T}_s].
	\vspace{-2mm}
\end{align}
In \eqref{eq:ASR-1}, $\Pr[t_a \leq t_{rd} | t_{rd} \geq \hat{T}_s]$ can be derived as 
\begin{equation}\label{eq:alpha-2}
\begin{split}
&\Pr[t_a \leq t_{rd} | t_{rd} \geq \hat{T}_s] 
=\Pr[t_{rd} \ge t_Q + \hat{T}_s | t_{rd} \ge \hat{T}_s] \\
&=\int_{0}^{\infty}\left\{ \int_{ x + \hat{T}_s}^{\infty}\! f_{t_{rd}}(y)\, \mathrm{d}y \right\} f_{t_Q}(x)\,\mathrm{d}x\\
&=\int_{0}^{\infty}\! e^{-\lambda_{rd}(x +\hat{T}_s)}\! f_{t_Q}(x)\,\mathrm{d}x 
=e^{-\lambda_{rd}\hat{T}_s}\!   \int_{0}^{\infty}\! e^{-\lambda_{rd}x} \!  f_{t_Q}(x)\,\mathrm{d}x\\
&=e^{-\lambda_{rd}\hat{T}_s} \mathcal{L}_{t_Q}(\lambda_{rd}) 
\end{split}
\end{equation}
where $\mathcal{L}_{t_Q}(s)$ is the Laplace transform of $t_{Q}$.
Since the interarrival time of the target devices is exponential distributed with mean $1/\lambda_t$, from M/G/1 queueing analysis in~\cite{Queueing_Kleinrock}, $\mathcal{L}_{t_Q}(\lambda_{rd})$ is given by
\begin{equation}\label{eq:alpha-3}
\mathcal{L}_{t_Q}(\lambda_{rd})  = \frac{(1-\rho)\lambda_{rd}}{\lambda_{rd}-\lambda_t+\lambda_t 
\mathcal{L}_{t_s}(\lambda_{rd})}
\end{equation} 
where $\rho=\lambda_{t}\hat{T}_{s}$. Since the authentication processing time $t_s$ is constant as $\hat{T}_s$, its Laplace transform is given by 
%
\begin{equation}\label{eq:alpha-4}
\begin{split}
\mathcal{L}_{t_s}(\lambda_{rd})
& =\int_{t=0}^{\infty}\!e^{-\lambda_{rd}t}f_{t_s}(t)\,\mathrm{d}t 
=e^{-\lambda_{rd}\hat{T}_s}.
\end{split}
\end{equation}
From (\ref{eq:alpha-2}), (\ref{eq:alpha-3}), and (\ref{eq:alpha-4}), we have 
\begin{equation}\label{eq:alpha-5}
\begin{split}
\Pr[t_a \leq t_{rd} | t_{rd} \geq \hat{T}_s]=
\frac{e^{-\lambda_{rd}\hat{T}_s} (1-\rho)\lambda_{rd}}{\lambda_{rd}-\lambda_{t}(1-e^{-\lambda_{rd}\hat{T}_s})}.
\end{split}
\end{equation}
The probability that the residence time of D2D communication is equal to or greater than $\hat{T}_s$ is given by 
 \begin{equation}\label{eq:alpha-6}
 \begin{split}
\Pr[t_{rd} \geq \hat{T}_{s}] 
 &=\int_{\hat{T_s}}^{\infty}\!\lambda_{rd}e^{-\lambda_{rd}x}\,\mathrm{d}x
 =e^{-\lambda_{rd}\hat{T}_s}.
 \end{split} 
 \end{equation}
By (\ref{eq:alpha-5}) and (\ref{eq:alpha-6}), we finally obtain \eqref{eq:ASR}. 
\end{IEEEproof}

Figure~\ref{fig:ana_sim_results} shows the authentication success rate with various $t_{rd}$ with time unit, $\hat{T}_s$. 
{In this figure, we can first see that our analysis on ASR is similar to the simulation result, which verifies the correctness of the analysis. 
We then also see that the ASR is more than $0.8$ when $t_{rd}$ is 10 times of $t_{a}$, which means that lower authentication processing time results in higher ASR regarding fixed residence time in D2D communication.
}

From~(\ref{eq:ASR}), we can explore the effects of $\lambda_{rd}$ and $\lambda_{t}$ on the ASR. Let the mean residence time of device in D2D communication be $1/\lambda_{rd}=c_{rd}\hat{T}_s$ and the mean interarrival time of device be $1/\lambda_{t}=c_{t}\hat{T}_s$, where $c_{rd},c_t\in[0,\infty]$. Thus, the ASR is represented as
\begin{equation}\label{eq:alpha-8}
\ASR = \frac{e^{-\frac{2}{c_{rd}}}\left(\frac{1}{c_{t}}-1\right)}{\frac{1}{c_{t}}-\frac{1}{c_{rd}}\left(1-e^{-\frac{1}{c_{rd}}}\right)}.
\end{equation}

{We show the ASR as a function of $c_{rd}$ for a given $c_t=2$. We have that if the system wants to obtain an ASR of at least 80\%, $c_{rd}$ should be greater than or equal to $11.091$, which is also confirmed by (\ref{eq:alpha-8}).} When the mean interarrival time of device is 2 times of the authentication processing time, the mean residence time of device in D2D communication should be greater than or equal to $11.091$ times of $\hat{T}_s$. With $\hat{T}_s=6.826$ ms~(\# of group is 50) as in table~\ref{table:computation_cost_GD2C} a device can achieve the ASR of more than 80\% by serving 146.49 authentication requests per second for $1/\lambda_t \geq 2\hat{T}_s$  where the mean residence time of device in D2D communication is greater than or equal to $75.707$ ms. 
This proves that the proposed NA-GD2C protocol is feasible to general D2D communications 
considering the density of D2D communication users and their mobility~\cite{D2D_LLAH15}. 


\subsubsection{ASR of CN-GD2C Protocool}


For the ASR in network-covered D2D communication, it needs to additionally consider the residence time that one of D2D devices is in the coverage of eNB. In figure~\ref{fig:d2d_auth_cn}, $\UE_i$ and $\UE_j$ negotiate the parameters for D2D communication and $\UE_i$ sends a request and its identity to start the authentication. Hence, both $\UE_i$ and $\UE_j$ need to reside in the coverage of each other before authentication. 
Thus, the ASR of CN-GD2C can be derived as follows.
\begin{lemma}\label{lemma:asr_CN-GD2C}
The authentication success rate of the proposed CN-GD2C protocol is given by 
\vspace{-2mm}
\begin{align}\label{eq:ASRCN}	
	\ASR \! = \! 
		\frac{
			e^{-2\hat{T}_s\left(\lambda_{r}\! +\! \lambda_{rd}\right)} \!  \left(1\! -\! \lambda_{t}\hat{T}_{s}\! \right)^2 \lambda_{rd}\lambda_{r}
		}{
			\left\{\! \lambda_{rd} \! -\!  \lambda_{t}\! \left(1\! -\! e^{-\lambda_{rd}\hat{T}_s}\! \right)\! \right\}
			\left\{\! \lambda_{r} \! - \! \lambda_{t}\! \left(1\! -\! e^{-\lambda_{r}\hat{T}_s}\! \right)\! \right\}
		}.
\end{align}
%
%
\end{lemma}

\begin{IEEEproof}
A successful authentication in CN-GD2C protocol requires that the residence times in D2D communication and eNB should be greater than or equal to the system authentication, including the waiting time in queue and the authentication processing time, for each device. Hence, the ASR can be represented as 
\begin{equation}\label{eq:asr_nc_1}
\begin{split}
\ASR &= \Pr \left[ t_{a} \leq t_{rd}, t_{a} \leq t_{r}  \right] 
= \Pr\left[ t_a \leq t_{rd}\right] \Pr\left[ t_{a} \leq t_{r} \right]
\end{split}
\end{equation}
where $\Pr[t_a \geq t_{r}]$ is given by 
\begin{equation}\label{eq:asr_nc_2}
\begin{split}
\Pr[t_a \leq t_{r} | t_{r} \geq \hat{T}_s] 
& =\Pr[t_Q+\hat{T}_s \leq t_{r} | t_{r} \geq T_s]\\
&=\Pr[t_{Q}\leq t_{r}-\hat{T}_s | t_{r} - \hat{T}_s \geq 0]\\
&=\frac{e^{-2\lambda_{r}\hat{T}_s} (1-\rho)\lambda_{r}}{\lambda_{r}-\lambda_{t}(1-e^{-\lambda_{r}\hat{T}_s})}.
\end{split}
\end{equation}
which is obtained in similar way as (\ref{eq:alpha-2}), (\ref{eq:alpha-3}), and (\ref{eq:alpha-4}). From (\ref{eq:ASR}), (\ref{eq:asr_nc_1}) and (\ref{eq:asr_nc_2}), we finally obtain
\begin{equation} \nonumber
\begin{split}
\ASR 
&= \frac{e^{-2\lambda_{rd}\hat{T}_s} (1-\rho)\lambda_{rd}}{\lambda_{rd}-\lambda_{t}(1-e^{-\lambda_{rd}\hat{T}_s})}\times
 \frac{e^{-2\lambda_{r}\hat{T}_s}(1-\rho)\lambda_{r}}{\lambda_{r}-\lambda_{t}(1-e^{-\lambda_{r}\hat{T}_s})}
\end{split}
\end{equation}
which is equal to \eqref{eq:ASRCN}.
\end{IEEEproof}


\begin{figure}[!t] 
\centering{
\psfrag{x}[tc][bc][0.6]{$c_{r}$~(unit:$\hat{T}_s$)~(CN-GD2C), $c_{rd}$~(unit:$\hat{T}_s$)~(NA-GD2C)}
		   \psfrag{y}[bc][tc][0.7]{Authentication success rate}
			\psfrag{na-analxxxxxxxxxxxxxxxxxxxxx}[Bl][Bl][0.6]{Analysis~(NA-GD2C)}
			\psfrag{na-simxxxxxxxxxxxxxxxxxxxxx}[Bl][Bl][0.6]{Simulation~(NA-GD2C)}
			\psfrag{cn-analxxxxxxxxxxxxxxxxxxxxx}[Bl][Bl][0.6]{Analysis~(CN-GD2C)}
			\psfrag{cn-simxxxxxxxxxxxxxxxxxxxxx}[Bl][Bl][0.6]{Simulation~(CN-GD2C)}
			\includegraphics[width=.35\textwidth,keepaspectratio]{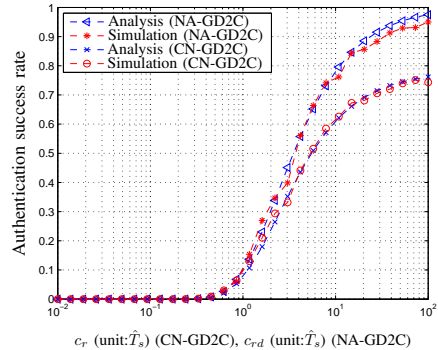}
}
	\caption{Authentication success rate of CN-GD2C protocol as a function of the ratio of the mean residence time in eNB to $\hat{T}_s$ (i.e., $c_{r} = \mbb{E}[t_r]/\hat{T}_s=\frac{1}{\lambda_{t}\hat{T}_s}$) with fixed $c_t=2$ and $c_{rd}=10$. Authentication success rate of NA-GD2C protocol as a function of the ratio of the mean residence time of D2D communication to the authentication processing time (i.e., $c_{rd} = \mathbb{E}[t_{rd}]/ \hat{T}_s = \frac{1}{\lambda_{rd}\hat{T}_s}$) with a fixed $c_t=2$.}
	\label{fig:ana_sim_results}
\end{figure}

Figure~\ref{fig:ana_sim_results} shows the analytic and simulation results of the ASR for network-covered D2D communications according to $t_r$ with time unit, $\hat{T}_s$. 
In this figure, we can also see that our analysis on ASR is similar to the simulation result, which verifies the correctness of the analysis.

By equation~(\ref{eq:ASRCN}), let the mean residence time of device in D2D communication be $c_{rd}\hat{T}_s$, the mean interarrival time of device be $c_{t}\hat{T}_s$, and the mean residence time of device in eNB be $c_{r}\hat{T}_s$, such that one can estimate the ASR as 
\begin{equation}\label{eq:asr_nc_4}
\begin{split}
\ASR \!&=\! 
	\frac{e^{-2\left(\frac{1}{c_{rd}}+\frac{1}{c_{r}}\right)} \!\left(1-\frac{1}{c_{t}}\right)\! \frac{1}{c_{rd}}}{\frac{1}{c_{rd}}-\frac{1}{c_t}\!\left(1-e^{-\frac{1}{c_{rd}}}\right)}
\! \times \!
\frac{(1-\frac{1}{c_t}) \frac{1}{c_r}}{\frac{1}{c_r}-\frac{1}{c_t}\left(1-e^{-\frac{1}{c_r}}\right)}.
\end{split}
\end{equation}
By equation~(\ref{eq:asr_nc_4}), there are various combinations of $\frac{1}{\lambda_{rd}}$ and $\frac{1}{\lambda_{r}}$ with fixed $\frac{1}{\lambda_t} = 2\hat{T}_s$ that make $\ASR'$ greater than or equal to 80\%. Considering that the range of the communication between devices is stricter than that between eNB and devices, we may select a combination to minimize $\frac{1}{\lambda_{rd}}$ guarantee the ASR can be greater than or equal to 80\%. To this end, a combination that $\frac{1}{\lambda_r}=12.915\hat{T}_s$, $\frac{1}{\lambda_{rd}}=83.022\hat{T}_s$, and $\frac{1}{\lambda_t}\geq 2\hat{T}_s$, guarantees the ASR is greater than or equal to 80\%. For $\hat{T}_s = 1.5716$ ms as in table~\ref{table:computation_cost_GD2C}, a device can serve at most 636.29 devices per second with $1/\lambda_t = 3.1432$ ms. Hence, the proposed CN-GD2C protocol completely satisfy the requirements of D2D communications.

From the above results, we can see that both NA-GD2C and CN-GD2C protocols are adequate for D2D communication regarding performance. 
\vspace{-2mm}
\section{Conclusions}\label{sec:conclusions}
\vspace{-1.5mm}

This work proposes two group AKE protocol for securing network-covered and network-absent D2D communications with traceability and guaranteeing end-to-end confidentiality {to network operators} for D2D applications.
Specifically, their security can be proven based on the security of pseudorandom function, pseudorandom permutation, IND-CCA IBE, linear encryption, and key-private encryption.
Furthermore, the performance analyses show the effect of the computation costs depending on ASRs by using M/G/1 queueing model, which 
reflects the effects of the residence times in D2D communication and in eNB, the inter-arrival time of devices, and the authentication processing time on the ASR. The communication costs of both protocol are considerably low even the new security properties are provided. 
Hence, the feasibility and scalability of this work are demonstrated to support efficient secure D2D communication in mobile networks. 
\vspace{-2mm}
\bibliographystyle{IEEEtran}
\bibliography{Anonymous_D2D_Rev_Richard_V9_toArXiv_}  
\end{document}